%% file: content/main.tex
\documentclass[pra,floatfix,superscriptaddress,notitlepage,longbibliography]{revtex4-1}

\usepackage{hyperref}
\usepackage[utf8]{inputenc}
\usepackage[UKenglish]{babel}
\usepackage{graphicx}
\graphicspath{ {Figures/} } 
\usepackage{dcolumn}
\usepackage{bm}
\usepackage{bbold}
\usepackage{xfrac}
\usepackage[dvipsnames]{xcolor}
\usepackage[ruled,vlined,linesnumbered]{algorithm2e}
\usepackage{setspace}
\usepackage[noend]{algpseudocode}
\usepackage{ragged2e}
\usepackage{amsmath}

\usepackage{circuitikz}
\usepackage{tikz}
\usetikzlibrary{shapes.multipart}
\usetikzlibrary{arrows, shapes.gates.logic.US, calc}
\usetikzlibrary{arrows,decorations.pathreplacing}
\usetikzlibrary{backgrounds,fit,decorations.pathreplacing,calc}
\usetikzlibrary{decorations.markings}
\usetikzlibrary{arrows.meta}

\newcommand{\bra}[1]{\ensuremath{\left\langle#1\right|}}
\newcommand{\ket}[1]{\ensuremath{\left|#1\right\rangle}}
\newcommand{\braket}[2]{\ensuremath{\left\langle#1\right|\!\!\left.#2\right\rangle}}
\newcommand{\ketbra}[2]{\ensuremath{\left|#2\right\rangle\!\!\left\langle#1\right|}}

\usepackage[normalem]{ulem}
\usepackage{color, graphpap}% Include figure files
\usepackage{enumitem}
\usepackage{amssymb}
\usepackage{amsthm}
\usepackage{subcaption}
\usepackage{import}
\usepackage{verbatim}
\usepackage{mathtools}
\usepackage{titlesec}
\usepackage{float}

\long\def\ca#1\cb{} %Use for commenting out: \ca...\cb

\renewcommand{\geq}{\geqslant}
\renewcommand{\leq}{\leqslant}

{}%         Body font
{}%         Indent amount (empty = no indent, \parindent 
\usepackage[textsize=footnotesize,backgroundcolor=red!25]{todonotes}

\begin{document}
\title{Lazy Quantum Walks with Native Multiqubit Gates}

\author{Steph Foulds}
\affiliation{Department of Physics, University of Strathclyde, Glasgow G4 0NG, UK}
\email{s.foulds@strath.ac.uk}

\author{Viv Kendon}
\affiliation{Department of Physics, University of Strathclyde, Glasgow G4 0NG, UK}

\begin{abstract} 
Quantum walks, the quantum analogue of the classical random walk, have been shown to underpin quantum algorithms for fluid dynamics. We propose the quantum half-adder gate method for quantum walks as a good benchmark algorithm, specifically to compare native two-qubit gate and native multiqubit gate implementations. Neutral atom hardware is a promising choice of platform for implementing quantum walks due to its ability to implement native multiqubit ($\geq\!3$-qubit) gates and to dynamically re-arrange qubits. Using detailed realistic error modelling for multiqubit Rydberg gates via two-photon adiabatic rapid passage, we present the gate sequences and predicted final state fidelities for some small one dimensional quantum walks, including lazy quantum walks; lazy quantum walks include a rest state, which is needed for quantum walks for fluid simulation. Our simulations pinpoint the sweet spot where native multiqubit gates provide an advantage compared with decomposing the gate into multiple smaller higher fidelity gates.
\end{abstract}
\maketitle

\import{content/}{intro.tex}
\import{content/}{background.tex}

\import{content/}{method.tex}
\import{content/}{results.tex}
\import{content/}{conc.tex}

\section*{Software}
Source code available at \href{https://github.com/sfoulds/multiqubit\_gate\_QWs}{https://github.com/sfoulds/multiqubit\_gate\_QWs}.

\begin{acknowledgments}
Thank you to Kathryn McInroy, Gerard Pelegrí, and Lara Janiurek for interesting discussions and vital expertise. Thank you additionally to Kathryn Lund and Lara Janiurek for careful and detailed review of the manuscript.
SF and VK funded by UKRI EPSRC grants EP/T026715/2, EP/T001062/1, EP/W00772X/2, EP/Y004566/1, EP/Z53318X/1, the UKRI DRI and STFC funded CCP-QC Bridge Project -- extended case studies for neutral atom hardware, and InnovateUK project number 10150308 SQALE2 TS/Z020964/1.
\end{acknowledgments}

\bibliography{content/main.bib}

\import{content/}{appendix.tex}

\end{document}

%% file: content/intro.tex
\section{Introduction}

%Engineering, astrophysics, and thermodynamics are among the many fields that employ fluid dynamics simulations. Useful problems are usually impossible to solve analytically, so classical solvers are used to simulate the partial differential equations via discrete updates. This requires the relevant nonlinear algebraic equations to be represented and solved on computers. Such simulations consume a significant proportion of high performance computing resources, and are thus an important potential application for quantum computing.

Discrete-time quantum walks (QWs) are the quantum analogue of the classical random walk: the movement of a walker on a lattice is dictated by the flip of a quantum `coin' at each step \cite{Aharonov_93,ambainis2004}. Due to the inference effects, quantum walks can exhibit exponentially faster hitting times \cite{Childs_2002, kempe2002} and algorithmic speed-up \cite{Childs_2003, shenvi_2003} compared to corresponding classical algorithms. The Dirac equation can be obtained from the continuum limit of a $d$-dimensional QW \cite{Strauch_2006,Strauch_2007,Arrighi_2014}. More recently, QWs have been shown to map to computational methods for fluid simulation such as the lattice Boltzmann method \cite{Succi_2015}, quantum fluid dynamics \cite{Hatifi_2019}, and smoothed-particle hydrodynamics \cite{Au_Yeung_2025} -- however in order to model a rest state with particle velocity zero in fluid simulation, the standard quantum walk must be extended to a \emph{lazy} quantum walk \cite{Childs2010}, performed on graphs with self loops. Lazy quantum walks have also been shown decrease search time compared to standard quantum walks \cite{Wong_2018}.
%
% Lara: You can cite the Childs 2009 paper as the first formulation of a lazy quantum walk which was used to prove that continuous QWs are the limit of discrete QWs. 
% Wong (2017), Faster Search by Lackadaisical Quantum Walks - cite this to show that lazy walks are usefuk in another context 
%
%Discrete-time quantum walks can be efficiently algorithmically implemented \cite{Douglas_2009} via quantum half-adder \cite{nielsen2002quantum} gate sequences, as employed in this work; such circuits require gates that act on at least three qubits, increasing with system size. We therefore believe the algorithm in this form to be a good benchmark: it is simple, transparent, relevant to quantum application, and exploits qubit connectivity.
%
One-dimensional quantum walks without a rest state have been implemented on a superconducting IBM device \cite{Wadhia_2024}, a trapped-ion quantum processor \cite{Huerta_Alderete_2020}, and a nuclear magnetic resonance (NMR) quantum processor \cite{Ryan_2005}. Lazy quantum walks, despite their theoretical utility, have not to our knowledge been implemented experimentally. 

Discrete-time quantum walks can be efficiently algorithmically implemented \cite{Douglas_2009} via quantum half-adder \cite{nielsen2002quantum} gate sequences, as employed in this work; such circuits require gates that act on at least three qubits, increasing with system size.
We believe the algorithm in this form to be a useful algorithmic benchmark that encompasses qubit connectivity, native/effective multiqubit gate fidelity, qubit coherence times, and state preparation and measurement (SPAM); it additionally has the benefits of being applicable in quantum computation and of simple construction.

Neutral atom quantum processors present a viable route to scalable quantum computing due to having dynamically reconfigurable qubit arrays \cite{Bluvstein_2022}, native multiqubit gates \cite{Levine_2019, Pelegri_2022}, high quality identical qubits \cite{Huft_2022, manetsch_2025}, and high fidelity single-qubit and two-qubit gates \cite{Levine_2019,Nikolov2023,Evered_2023}.
%
%Native multiqubit gates, achieved via the strong long-range interactions between highly excited Rydberg states \cite{Levine_2019, Pelegri_2022}, and mid-circuit qubit rearrangement are highly beneficial to efficiently implementing quantum algorithms requiring high connectivity \cite{mcinroy_2024,Wagner_2024}.
%
Previous benchmarks applied to neutral atom hardware include randomised benchmarking \cite{xia_2015, Nikolov2023}, quantum volume metrics \cite{mcinroy_2024}, and
algorithmic benchmarks
\cite{Wagner_2024, mcinroy_2024, rava_2025}.
In particular, \citeauthor{Wagner_2024} \cite{Wagner_2024} simulate circuits from the the Quantum Economic Development Consortium (QED-C) suite of algorithmic benchmarks \cite{Lubinski_2023} with a neutral atom noise model, in order to compare the final state fidelities obtained when utilising all-to-all connectivity as opposed to only nearest neighbour connectivity. All-to-all connectivity can be achieved in neutral atom processors via mid-circuit movement of qubits, thereby reducing gate depth. The authors find that all-to-all connectivity improves final state fidelity by 10\% – 15\%.
\citeauthor{mcinroy_2024} \cite{mcinroy_2024}, as well as finding the quantum volume of a 9-qubit neutral atom device to be competitive with a superconducting device of the same size, use Grover's search algorithm \cite{nielsen2002quantum} to compare success probabilities with and without utilising the multiqubit CCZ gate. With the native CCZ gate, the authors report a predicted 43\% increase in success probability without loss-correction and a 21\% increase with loss-correction.
We extend this analysis to quantum walks as a suitable benchmark and by the inclusion in our model of the as yet unrealised native C3Z gate, as described by \citeauthor{Pelegri_2022} \cite{Pelegri_2022}.
%We aim to extend the present analysis by highlighting not only the utility of neutral atom hardware's dynamically reconfigurable qubit array, but also its ability to perform native multiqubit gates.
%
%Constructing a metric by which to compare various quantum hardwares which is both deterministic and useful is a complicated task, with each benchmark prioritising various features.

%
Using error modelling for native 2-, 3-, and 4-qubit Rydberg gates via two-photon adiabatic rapid passage \cite{Pelegri_2022}, SPAM, and passive error \cite{Bluvstein_2022}, we present the gate sequences and final state fidelities for quantum walks with and without a rest state on 4- to 16-node rings. 
Our results highlight the utility of the native multiqubit gates and mid-circuit array programmability in the neutral atom platform and can guide the progress of neutral atom hardware with potential applications. Our detailed comparisons of multiqubit gates with decompositions into longer but higher fidelity gate sequences show that, for current error rates, three and four qubit gates provide a significant advantage. Composite fidelity results for five or more qubit gates suggest this would deliver only marginal further gains.

The paper is organised as follows.  In section \ref{sec:back} we provide brief background on quantum computation via gates and quantum walks.  In section \ref{sec:method} we describe the gate sequences and neutral atom models we use.  In section \ref{sec:results} we present the results of our simulations, and in section \ref{sec:conc} we summarise and conclude. 

%To this end, in this work we simulate with near-term and further-term neutral atom error modelling some quantum walks on small, one-dimensional rings with and without rest states.

%% file: content/background.tex
\section{Background}\label{sec:back}

\subsection{Qubits, gates, and fidelities}
In this work we exclusively use \emph{pure} 2-dimensional qubits as inputs, as this is a good approximation for near-term neutral atom qubits (see Section \ref{neutral-atoms}). A pure one-qubit is represented by state vector \cite{nielsen2002quantum}
\begin{align}
    \ket{\psi_1} = A_0 \ket{0} + A_1 \ket{1} = \begin{pmatrix}
        A_0 \\ A_1
    \end{pmatrix}
\end{align}
and a pure $n$-qubit composite state by state vector
\begin{align}
    \ket{\psi_n} = \sum_{\sigma\in\{0,1\}^{\otimes n}} A_\sigma \ket{\sigma} = \begin{pmatrix}
        A_{0^{\otimes n}} \\ \vdots \\ A_{1^{\otimes n}}
    \end{pmatrix}
\end{align}
where the probability $P(\sigma)$ of measuring bit string $\sigma$ in the computational basis is related to its amplitude $A_\sigma$ with $P(\sigma) = |\braket{\sigma}{\psi}|^2= |A_\sigma|^2$ where $\braket{\sigma}{\sigma'}=\delta_{\sigma\sigma'}$ (and $\delta_{ij}$ is the Kronecker delta). However, due to the various errors acting on the qubits, output states may be mixed qubit states represented by $\rho_A = \text{tr}_{B} (\ketbra{\psi}{\psi}_{AB}) = \sum_{j \in \{0,1\}^{\text{len}(B)}} \bra{j}_B \ketbra{\psi}{\psi}_{AB} \ket{\psi}_B$.

We use the gate model of quantum computation, where in the ideal case unitary gate operators U are applied to a pure state $\ket{\psi}$ with $\ket{\psi'}=U\ket{\psi}$. The gates used in this work, in their ideal unitary form, are as follows: \cite{nielsen2002quantum}
\begin{itemize}
    \item Rotation about the y-axis $\text{R}_y\left(\theta \right)$ of a single qubit state:
        \begin{align}
            \text{R}_y\left(\theta \right) =
            \begin{pmatrix}
                \cos\left(\frac{\theta}{2}\right) & -\sin\left(\frac{\theta}{2}\right)\\
                \sin\left(\frac{\theta}{2}\right) & \cos\left(\frac{\theta}{2}\right)
            \end{pmatrix} ;
        \end{align}
    \item The Pauli-X gate X, which `flips' a single qubit such that $\ket{0} \leftrightarrow \ket{1}$:
        \begin{align}
            \text{X} =
            \begin{pmatrix}
                0 & 1\\
                1 & 0
            \end{pmatrix} ;
        \end{align}
    \item The set of controlled NOT gates $\{ \text{C}k\text{X} | \,\, k \in \mathbb{N} \text{ and } k \geq 1\}$:
    \begin{align}
        \text{CX}_\text{ideal} &= \begin{pmatrix}
                1 & 0 & 0 & 0 \\
                0 & 1 & 0 & 0 \\
                0 & 0 & 0 & 1 \\
                0 & 0 & 1 & 0
        \end{pmatrix} ; \\
            \text{C}k\text{X}_\text{ideal} &= \ketbra{1^{\otimes k}0}{1^{\otimes k}1} + \ketbra{1^{\otimes k}1}{1^{\otimes k}0} + \sum_{\substack{\sigma_k \in \{0,1\}^{\otimes k},\\ \sigma_k \neq 1^{\otimes k}}}\sum_{\sigma_1 \in \{0,1\}} \ketbra{\sigma_k \sigma_1}{\sigma_k \sigma_1} \\
            &= \begin{pmatrix}
                1 & 0 \\
                0 & 1\\
                  &  & \ddots \\
                  &  && 0 & 1\\
                  &  && 1 & 0
            \end{pmatrix}
        \end{align}
    which acts on a $k+1$ qubit state, and flips ($\ket{0} \leftrightarrow \ket{1}$) the target qubit -- here the last qubit -- if the $k$ control qubits are in state $\ket{1}^{\otimes k}$;
    \item The set of controlled phase gates $\{ \text{C}k\text{Z} | \,\, k \in \mathbb{N} \text{ and } k \geq 1\}$:
        \begin{align}
            \text{CZ}_\text{ideal} &= \begin{pmatrix}
                1 & 0 & 0 & 0 \\
                0 & -1 & 0 & 0 \\
                0 & 0 & -1 & 0 \\
                0 & 0 & 0 & -1
        \end{pmatrix} ;\\
            \text{C}k\text{Z}_\text{ideal} &= \ket{0}^{\otimes (k+1)}\bra{0}^{\otimes (k+1)} - \sum_{\substack{\sigma \in \{0,1\}^{\otimes (k+1)},\\ \sigma \neq 0^{\otimes(k+1)}}} \ketbra{\sigma}{\sigma} \\
            &= \text{diag}(1, -1, ..., -1)
        \end{align}
    which changes the sign of the amplitude of the $(k+1)$-qubit state ($A_\sigma \leftrightarrow -A_\sigma$) it is applied to so long as said qubits are not in state $\ket{0}^{\otimes (k+1)}$.
\end{itemize}

In order to model realistic lossy gates, we use non-unitary effective matrices in the place of unitary gates. These effective gates $\text{U}_\text{eff}$ are characterised by their gate fidelity $\mathcal{F}$ compared to their corresponding ideal unitary gate $\text{U}_\text{ideal}$, defined with the limited tomography prescription \cite{Pelegri_2022}:
\begin{align}
    \mathcal{F}(\text{U}_\text{eff}) = |\bra{\phi}\text{U}_\text{eff}^\dagger \text{U}_\text{ideal} \ket{\phi} |^2
\end{align}
where equal superposition state $\ket{\phi}=\frac{1}{\sqrt{2^m}}\sum_{s \in \{0,1\}^{\otimes m}} \ket{s}$. Gate fidelity is bounded as $0 \leq \mathcal{F} \leq 1$: if $\mathcal{F}(\text{U}_\text{eff})=1$ then $\text{U}_\text{eff} \equiv \text{U}_\text{ideal}$; if $\mathcal{F}(\text{U}_\text{eff})=0$ then the states $\ket{\Psi}=\text{U}_\text{ideal}\ket{\psi}$ and $\ket{\Phi}=\text{U}_\text{eff}\ket{\psi}$ are orthogonal.

In order to determine the resulting accuracy of an algorithm run with realistic error modelling, we calculate the state fidelity of the final state obtained with the final state with an errorless model.
We define state fidelity in terms of the Hellinger distance \cite{Hellinger, IBM_fid} between two probability distributions $p$ and $q$:
\begin{align}
    H^2 &= \frac{1}{2} \sum_z \left(\sqrt{p(z)} - \sqrt{q(z)} \right)^2 \\
    &= \frac{1}{2}\left(\sum_zp(z) + \sum_zq(z)\right) - \braket{\sqrt{p}}{\sqrt{q}} \nonumber
\end{align}
therefore we define our state fidelity \cite{IBM_fid} as
\begin{align}
    f &= ( 1 - H^2)^2 \\
    &= \left( 1 - \frac{1}{2} \sum_z \left(\sqrt{P_\text{ideal}(z)} - \sqrt{P_\text{eff}(z)} \right)^2 \right)^2 \\
    &= \left( \frac{1}{2} + \frac{1}{2} \sum_z \left(2\sqrt{P_\text{ideal}(z) P_\text{eff}(z)} - P_\text{eff}(z) \right) \right)^2 \\
    &= \left( \frac{1}{2}\left(1 - \sum_z P_\text{eff}(z)\right) + \braket{\sqrt{P_\text{ideal}}}{\sqrt{P_\text{eff}}} \right)^2
\end{align}
where $P_\text{eff}(z)$ is the probability of measuring $z$ in a model with errors, and $P_\text{ideal}(z)$ is the probability of measuring $z$ for an errorless model. Note that while $\sum_z P_\text{ideal}(z) = 1$, we do not renormalise during our simulation in order to model qubit population loss, therefore $\sum_z P_\text{eff}(z) < 1$.
The state fidelity is bounded as $0 \leq f \leq 1$ where $f=1$ denotes complete agreement between the two probability sets and $f=0$ denotes approximately orthogonal probability distributions, as if $f=0 \rightarrow |\braket{\sqrt{P_\text{ideal}}}{\sqrt{P_\text{eff}}}| \equiv |\sum_z\sqrt{P_\text{ideal}(z) P_\text{eff}(z)}| = \frac{1}{2}(1- \sum_z P_\text{eff}(z)) \ll 1$. 
We have chosen this measure in order to make a comparison with the results obtained in Ref \cite{Wadhia_2024}.

\subsection{Quantum walks} \label{sec:BgQWs}
In this work we consider one dimensional discrete time quantum walks on finite line segments or cycles \cite{Aharonov_93, Aharanov-2001}. 
The `walker' traverses an $N$-node line with position labels $\{x_j\}^N_{j=0}$%$x \in \mathbb{Z}_N$
. In a quantum computer, we encode the position label into the qubit register using $n = \log_2 N$ qubits. 
%In this work we binary encode the position state with $n$ qubits s.t. $x = 0 \rightarrow \ket{x} = \ket{0}^{\otimes n}$ ... $x = N-1 \rightarrow \ket{x} = \ket{1}^{\otimes n}$. 
Therefore, no physical walk is taking place, only an encoded one. Efficient quantum walks must be encoded in this way so that the position qubit requirement scales  logarithmically with graph
size instead of polynomially for physical walks \cite{Douglas_2009}. We use cyclic boundary conditions such that $x_N$ is one step from $x_0$ in the negative direction, so that the line becomes a ring/cycle graph.

Each step $t$ of the walk, the coin is `tossed' with coin operator C, and then the shift operator S moves the walker clockwise or anticlockwise on the line according to the coin state. For a standard one-qubit coin (1q-coin) QW, the state of the coin at step $t$ is $\ket{c(t)} = A_0(t) \ket{0} + A_1(t) \ket{1}$. The walker state at step $t$ is then $\ket{\psi(t)}=\sum_{x,c} A_{x,c}(t)\ket{x, c}$ with Hilbert space 
\(\mathcal{H}_{xc} = \mathcal{H}_x \otimes \mathcal{H}_c\). The walker state evolves each step with $\ket{\psi(t+1)}=\text{SC}_t\ket{\psi(t)}$, where 
%\cite{Kendon_2006}
\begin{align}
    \text{S} \ket{x, 0} &= \ket{x-1, 0} \\
    \text{S} \ket{x, 1} &= \ket{x+1, 1},
\end{align}
The coin operator defines the dynamic of the walk and can in theory take the form of any unitary acting on $\mathcal{H}_{xc}$, but will often only act on $\mathcal{H}_c$.

\begin{figure}[h!]
    \centering
    %\hspace{0.2cm}
    \begin{subfigure}{0.49\textwidth}
        \begin{tikzpicture}[thick]
            \tikzset{phase/.style = {draw,fill,shape=circle,minimum size=5pt,inner sep=0pt},phase2/.style = {draw,shape=circle,minimum size=5pt,inner sep=0pt}} \matrix[row sep=1cm, column sep=1cm] (circuit) {
            % G4
            \node[phase] (W41) {}; &\node[phase] (W42) {}; \\
            \node[phase] (W43) {}; &\node[phase] (W44) {}; \\
            }; \begin{pgfonlayer}{background}
            % G4
            \draw[-stealth, MidnightBlue,line width=0.05cm,line width=0.05cm] (W41) -- (W42); \draw[-stealth, MidnightBlue,line width=0.05cm,line width=0.05cm] (W42) -- (W41); \draw[-stealth, MidnightBlue,line width=0.05cm, line width=0.05cm] (W43) -- (W44); \draw[-stealth, MidnightBlue,line width=0.05cm] (W44) -- (W43);
            \draw[-stealth, MidnightBlue,line width=0.05cm,line width=0.05cm] (W41) -- (W43); \draw[-stealth, MidnightBlue,line width=0.05cm,line width=0.05cm] (W43) -- (W41); \draw[-stealth, MidnightBlue,line width=0.05cm,line width=0.05cm] (W42) -- (W44); \draw[-stealth, MidnightBlue,line width=0.05cm,line width=0.05cm] (W44) -- (W42); \draw[-stealth, Purple, densely dashed, line width=0.05cm,] (W41) to[out=100,in=170,loop] node[midway, above left, black] {0: $\ket{00}$} (); \draw[-stealth, Purple, densely dashed, line width=0.05cm,] (W42) to[out=10,in=80,loop] node[midway, above right, black] {1: $\ket{01}$} (); \draw[-stealth, Purple, densely dashed, line width=0.05cm,] (W43) to[out=-100,in=-170,loop] node[midway, below left, black] {3: $\ket{11}$} (); \draw[-stealth, Purple, densely dashed, line width=0.05cm,] (W44) to[out=-10,in=-80,loop] node[midway, below right, black] {2: $\ket{10}$} ();
             \end{pgfonlayer}
        \end{tikzpicture}
        \caption{Graph representation of a 4-node ring in state space.}
    \end{subfigure}
    %\hspace{-1cm}
    \begin{subfigure}{0.49\textwidth}
    %\centering
        \begin{tikzpicture}[line width=0.05cm]
            \tikzset{phase/.style = {draw,fill,shape=circle,minimum size=5pt,inner sep=0pt}}
            \matrix[row sep=0.05cm, column sep=1cm] (circuit) {
            \node (x1) {$x_1$}; &\node (x2) {$x_2$};
            \\
            \node [circle split,draw,rotate=90] (y) {\rotatebox{-90}{$0$}\nodepart{lower} \rotatebox{-90}{$1$}};
            &\node [circle split,draw,rotate=90] (y) {\rotatebox{-90}{$0$}\nodepart{lower} \rotatebox{-90}{$1$}};
            \\ \node (bleh) {}; \\
            \node (c1) {$c_1$}; &\node (c2) {$c_2$};
            \\
            \node [circle split,draw=MidnightBlue,rotate=90] (y) {\rotatebox{-90}{$0$}\nodepart{lower} \rotatebox{-90}{$1$}};
            &\node [circle split,draw=Purple,dashed,rotate=90] (y) {\rotatebox{-90}{$0$}\nodepart{lower} \rotatebox{-90}{$1$}};
            \\
            }; \begin{pgfonlayer}{background}
            \end{pgfonlayer}
        \end{tikzpicture}
        \caption{Physical qubits for a 4-node ring.}
    \end{subfigure}
    \\
    \vspace{0.5cm}
    \begin{subfigure}{0.49\textwidth}
        \begin{tikzpicture}[very thick]
            \tikzset{phase/.style = {draw,fill,shape=circle,minimum size=5pt,inner sep=0pt},phase2/.style = {draw,shape=circle,minimum size=5pt,inner sep=0pt}} \matrix[row sep=0.6cm, column sep=0.6cm] (circuit) {
            % G4
            &\node[phase] (W000) {};
            &\node[phase] (W001) {}; \\
            \node[phase] (W111) {};
            &&&\node[phase] (W010) {}; \\
            \node[phase] (W110) {};
            &&&\node[phase] (W011) {}; \\
            &\node[phase] (W101) {};
            &\node[phase] (W100) {}; \\
            }; \begin{pgfonlayer}{background}
            % G4
            \draw[-stealth, MidnightBlue,line width=0.05cm] (W000) -- (W001); \draw[-stealth, MidnightBlue,line width=0.05cm] (W001) -- (W000); \draw[-stealth, MidnightBlue,line width=0.05cm] (W001) -- (W010); \draw[-stealth, MidnightBlue,line width=0.05cm] (W010) -- (W001);
            \draw[-stealth, MidnightBlue,line width=0.05cm] (W010) -- (W011); \draw[-stealth, MidnightBlue,line width=0.05cm] (W011) -- (W010);
            \draw[-stealth, MidnightBlue,line width=0.05cm] (W011) -- (W100); \draw[-stealth, MidnightBlue,line width=0.05cm] (W100) -- (W011);
            \draw[-stealth, MidnightBlue,line width=0.05cm] (W100) -- (W101); \draw[-stealth, MidnightBlue,line width=0.05cm] (W101) -- (W100);
            \draw[-stealth, MidnightBlue,line width=0.05cm] (W101) -- (W110); \draw[-stealth, MidnightBlue,line width=0.05cm] (W110) -- (W101);
            \draw[-stealth, MidnightBlue,line width=0.05cm] (W110) -- (W111); \draw[-stealth, MidnightBlue,line width=0.05cm] (W111) -- (W110);\draw[-stealth, MidnightBlue,line width=0.05cm] (W000) -- (W111); \draw[-stealth, MidnightBlue,line width=0.05cm] (W111) -- (W000); \draw[-stealth, Purple, densely dashed, line width=0.05cm] (W000) to[out=100,in=170,loop] node[midway, above, black] {0: $\ket{000}$} ();
            \draw[-stealth, Purple, densely dashed, line width=0.05cm] (W001) to[out=10,in=80,loop] node[midway, above right, black] {1: $\ket{001}$} ();
            \draw[-stealth, Purple, densely dashed, line width=0.05cm] (W010) to[out=-10,in=60,loop] node[midway, above right, black] {2: $\ket{010}$} ();
            \draw[-stealth, Purple, densely dashed, line width=0.05cm] (W011) to[out=-60,in=10,loop] node[midway, below right, black] {3: $\ket{011}$} ();
            \draw[-stealth, Purple, densely dashed, line width=0.05cm] (W100) to[out=-10,in=-80,loop] node[midway, below right, black] {4: $\ket{100}$} ();
            \draw[-stealth, Purple, densely dashed, line width=0.05cm] (W101) to[out=-100,in=-170,loop] node[midway, below, black] {5: $\ket{101}$} ();
            \draw[-stealth, Purple, densely dashed, line width=0.05cm] (W110) to[out=-100,in=-170,loop] node[midway, below left, black] {6: $\ket{110}$} ();
            \draw[-stealth, Purple, densely dashed, line width=0.05cm] (W111) to[out=100,in=170,loop] node[midway, above left, black] {7: $\ket{111}$} ();
             \end{pgfonlayer}
            \end{tikzpicture}
            \caption{Graph representation of a 8-node ring in state space.}
        \end{subfigure}
        %\hspace{-1cm}
        \begin{subfigure}{0.49\textwidth}
            \begin{tikzpicture}[line width=0.05cm]
                \tikzset{phase/.style = {draw,fill,shape=circle,minimum size=5pt,inner sep=0pt}}
                \matrix[row sep=0.05cm, column sep=0pt] (circuit) {
                \node (x1) {$x_1$}; &&\node (x2) {$x_2$}; &&\node (x3) {$x_3$};
                \\
                \node [circle split,draw,rotate=90] (y) {\rotatebox{-90}{$0$}\nodepart{lower} \rotatebox{-90}{$1$}};
                &&\node [circle split,draw,rotate=90] (y) {\rotatebox{-90}{$0$}\nodepart{lower} \rotatebox{-90}{$1$}};
                &&\node [circle split,draw,rotate=90] (y) {\rotatebox{-90}{$0$}\nodepart{lower} \rotatebox{-90}{$1$}};
                \\ \node (bleh) {}; \\
                &\node (c1) {$c_1$}; &&\node (c2) {$c_2$};
                \\
                &\node [circle split,draw=MidnightBlue,rotate=90] (y) {\rotatebox{-90}{$0$}\nodepart{lower} \rotatebox{-90}{$1$}};
                &&\node [circle split,draw=Purple,dashed,rotate=90] (y) {\rotatebox{-90}{$0$}\nodepart{lower} \rotatebox{-90}{$1$}};
                \\
                \\
                }; \begin{pgfonlayer}{background}
                \end{pgfonlayer}
            \end{tikzpicture}
        \caption{Physical qubits for an 8-node ring.}
        \end{subfigure}
    \caption{Binary position and coin encodings for a $2^n$-node ring, with the $2^n$ position nodes on the left encoded by the $n$ physical qubits $x_j$ in black on the right. Arrows denote available shift directions around the ring for a 2-qubit coin/lazy QW: the solid blue clockwise and anticlockwise directions are encoded by the solid blue physical qubit $c_1$, and the dashed purple rest state is encoded by the dashed purple physical qubit $c_2$. Omission of $c_2$ and therefore the rest state yields a 1-qubit coin/standard QW.}
    \label{encodings}
\end{figure}

Quantum walks as the quantum counterpart to classical random walks, first conceptualised by Feynman (1985) \cite{Feynman_85} and first described by Aharonov \emph{et al.} (1993) \cite{Aharonov_93}. QWs were first discretised in time and space, as in this work, by Aharonov \emph{et al.} (2001) \cite{Aharanov-2001};
the latter work shows that the walker's position spreads faster during a quantum walk than during a classical random walk: proportional to the number of steps $t$ as opposed to the the square root of the number of steps, providing a quadratic speed up. Several algorithmic advantages with the use of QWs have since been identified, such as quadratic speed up when applied to the search of an unsorted database \cite{shenvi_2003} and exponential speed up to find routes across specific networks \cite{Childs_2003}. More generally, continuous and discrete QWs have been shown to be universal models of quantum computation (capable of simulating any other models) \cite{childs_uni, Lovett_2010} and therefore are capable (not necessarily efficiently) of computing anything computable by other universal quantum computers.

Recently, there has been interest in fluid simulation as a potential application for QWs, largely motivated by the natural mapping of the shift and coin structure of QWs to the stream and collide structure of many fluid simulation methods. 
Notably, discrete-time QWs have been shown to provide an algorithmic basis for the quantum lattice Boltzmann method \cite{Succi_2015, todorova_2020}, quantum fluid dynamics \cite{Hatifi_2019, Molfetta_fluid}, and smoothed particle hydrodynamics \cite{Au_Yeung_2025}. However, the lattice Boltzmann method requires zero in the available velocity set \cite{LB} such that particles can be at rest. In the QW setting, this maps to adding self loops to the graph on which the quantum walker walks. A quantum walk on such a graph is called a \emph{lazy} or \emph{lackadaisical} quantum walk, 
%A potential application for QWs is fluid simulation, as []. However, fluid simulation requires a rest state, during which the fluid is stationary. In the QW setting this maps to lazy quantum walks,
which is very similar to standard QWs except that the dimensionality of the coin state is increased to encompass a rest state.
Aside from their potentiation application in fluid simulation, lazy quantum walks have been shown to provide a logarithmic improvement compared to standard loopless quantum walks when used for search on a 2D grid of $N$ vertices \cite{Wong_2018}, with success probability near 1 in $O(\sqrt{N\log N})$ steps as opposed to $O(\sqrt{N}\log N)$ steps. Adding a self-loop to each vertex of a complete graph of $N$ vertices increases the success probability of Grover's search for a marked vertex from 1/2 to 1 \cite{Wong_2015,Giri_2019}.

Although increasing the dimensionality of the coin state can be done with a single qutrit (three-dimensional) coin, to stay in line with near-term hardware we instead encode the coin state $\ket{c}$ into two qubits with $c \in \{0,1\}^{\otimes 2}$. Whether the walker steps clockwise or anticlockwise is controlled on the first coin qubit, as with the standard QW. But in the 2-qubit coin QW whether the walker moves at all is controlled on the state of the second coin qubit; therefore the shift dynamics become:
\begin{align}
    \text{S} \ket{x, 00} &= \ket{x, 00} \\
    \text{S} \ket{x, 01} &= \ket{x-1, 01} \\
    \text{S} \ket{x, 10} &= \ket{x, 10} \\
    \text{S} \ket{x, 11} &= \ket{x+1, 11} .
\end{align}
This is illustrated in Figure \ref{encodings}, where the ring graphs $\{x_j\}^N_{j=0}$ with shift directions anticlockwise, clockwise, and rest are shown for $N=4$ and $N=8$. Alongside are the qubits required to encode these positions and shift directions.

\subsection{Neutral atom quantum computing}

Neutral atom computing is performed on a programmable array of identical neutral atoms in optical tweezer traps \cite{Huft_2022,manetsch_2025}. Its strengths lie in its high quality, long-life qubits \cite{manetsch_2025} and the strong long-range interactions between high energy Rydberg levels, resulting in the ability to perform native multiqubit control gates \cite{Levine_2019, Pelegri_2022}. Further, transport via optical tweezers allows for coherence-preserving qubit transport \cite{Bluvstein_2022, manetsch_2025}, unlocking all-to-all qubit connectivity.

Computation is performed in the hyperfine basis $\{\ket{0}, \ket{1}\}$ with a high energy Rydberg state $\ket{r}$ as an intermediary; $\ket{1}$ is coupled to $\ket{r}$ via a two-photon transition. Rydberg states experience dipole-induced pairwise interaction, and therefore within the blockade distance there can only be one Rydberg excitation \cite{Urban_2009}. Native multiqubit gates are controlled phase gates $\text{C}_k\text{Z}$, achieved with two ARP (adiabatic rapid passage) \cite{Pelegri_2022, Levine_2019} pulses and a global single-qubit rotation $\text{Z}(\phi)$. In theory the value of $k$ is only limited by the architecture of the processor's atoms array, however in practice only native CZ and CCZ gates have been implemented in a neutral atom setting \cite{Levine_2019, Evered_2023}.
\begin{figure}[h!]
    %%%% |00>
    \begin{tikzpicture}[
      scale=0.5,
      level/.style={thick},
      virtual/.style={thick,densely dashed},
      trans/.style={thick,<->,shorten >=2pt,shorten <=2pt,>=stealth},
      classical/.style={thin,double,<->,shorten >=4pt,shorten <=4pt,>=stealth}
    ]

    %%% 1
    %% A
    % Draw the energy levels.
    \draw[level] (2cm,4em) -- (0cm,4em) node[left] {\ket{r}} ;
    \draw[level] (2cm,-1em) -- (0cm,-1em) node[left] {\ket{1}};
    \draw[level] (2cm,-3em) -- (0cm,-3em) node[left] {\ket{0}};
    % qubit
    \draw [fill=blue] (1cm,-1.05cm) circle (.3cm);
    % pulse
    \draw[-stealth] (1cm,-1em) -- (1cm,4em);
    %% B
    % Draw the energy levels.
    \draw[level] (4.5cm,4em) -- (2.5cm,4em);
    \draw[level] (4.5cm,-1em) -- (2.5cm,-1em);
    \draw[level] (4.5cm,-3em) -- (2.5cm,-3em);
    % qubit
    \draw [fill=blue] (3.5cm,-1.05cm) circle (.3cm);
    
    %%% 2
    \draw[->] (4.5cm,8em) -- (6cm,8em) node[midway, left, xshift=-1.1cm] {\ket{00}};
    %% A
    % Draw the energy levels.
    \draw[level] (8cm,4em) -- (6cm,4em);
    \draw[level] (8cm,-1em) -- (6cm,-1em);
    \draw[level] (8cm,-3em) -- (6cm,-3em);
    % qubit
    \draw [fill=blue] (7cm,-1.05cm) circle (.3cm);
    %% B
    % Draw the energy levels.
    \draw[level] (10.5cm,4em) -- (8.5cm,4em);
    \draw[level] (10.5cm,-1em) -- (8.5cm,-1em);
    \draw[level] (10.5cm,-3em) -- (8.5cm,-3em);
    % qubit
    \draw [fill=blue] (9.5cm,-1.05cm) circle (.3cm);
    % pulse
    \draw[-stealth] (9.5cm,-1em) -- (9.5cm,4em);

    %%% 3
    \draw[->] (10.5cm,8em) -- (12cm,8em) node[midway, left, xshift=-1.1cm] {\ket{00}};
    %% A
    % Draw the energy levels.
    \draw[level] (14cm,4em) -- (12cm,4em);
    \draw[level] (14cm,-1em) -- (12cm,-1em);
    \draw[level] (14cm,-3em) -- (12cm,-3em);
    % qubit
    \draw [fill=blue] (13cm,-1.05cm) circle (.3cm);
    %% B
    % Draw the energy levels.
    \draw[level] (16.5cm,4em) -- (14.5cm,4em);
    \draw[level] (16.5cm,-1em) -- (14.5cm,-1em);
    \draw[level] (16.5cm,-3em) -- (14.5cm,-3em);
    % qubit
    \draw [fill=blue] (15.5cm,-1.05cm) circle (.3cm);
    % pulse
    \draw[-stealth] (15.5cm,4em) -- (15.5cm,-1em);

    %%% 4
    \draw[->] (16.5cm,8em) -- (18cm,8em) node[midway, left, xshift=-1.1cm] {\ket{00}};
    %% A
    % Draw the energy levels.
    \draw[level] (20cm,4em) -- (18cm,4em);
    \draw[level] (20cm,-1em) -- (18cm,-1em);
    \draw[level] (20cm,-3em) -- (18cm,-3em);
    % qubit
    \draw [fill=blue] (19cm,-1.05cm) circle (.3cm);
    %% B
    % Draw the energy levels.
    \draw[level] (22.5cm,4em) -- (20.5cm,4em);
    \draw[level] (22.5cm,-1em) -- (20.5cm,-1em);
    \draw[level] (22.5cm,-3em) -- (20.5cm,-3em);
    % qubit
    \draw [fill=blue] (21.5cm,-1.05cm) circle (.3cm);
    % pulse
    \draw[-stealth] (19cm,4em) -- (19cm,-1em);

    %%% 5
    \draw[->] (22.5cm,8em) -- (24cm,8em) node[midway, left, xshift=-1.1cm] {\ket{00}} node[midway, right, xshift=1.1cm] {\ket{00}};
    %% A
    % Draw the energy levels.
    \draw[level] (26cm,4em) -- (24cm,4em);
    \draw[level] (26cm,-1em) -- (24cm,-1em);
    \draw[level] (26cm,-3em) -- (24cm,-3em);
    % qubit
    \draw [fill=blue] (25cm,-1.05cm) circle (.3cm);
    %% B
    % Draw the energy levels.
    \draw[level] (28.5cm,4em) -- (26.5cm,4em);
    \draw[level] (28.5cm,-1em) -- (26.5cm,-1em);
    \draw[level] (28.5cm,-3em) -- (26.5cm,-3em);
    % qubit
    \draw [fill=blue] (27.5cm,-1.05cm) circle (.3cm);

    \draw[-] (-1.5cm, -5em) -- (29cm, -5em);    
    \end{tikzpicture}

    %%%% |01>
    \begin{tikzpicture}[
      scale=0.5,
      level/.style={thick},
      virtual/.style={thick,densely dashed},
      trans/.style={thick,<->,shorten >=2pt,shorten <=2pt,>=stealth},
      classical/.style={thin,double,<->,shorten >=4pt,shorten <=4pt,>=stealth}
    ]
    \tikzset{phase/.style = {draw,fill=cyan,shape=circle,minimum size=20pt,inner sep=0pt},}

    %%% 1
    %% A
    % Draw the energy levels.
    \draw[level] (2cm,4em) -- (0cm,4em) node[left] {\ket{r}} ;
    \draw[level] (2cm,-1em) -- (0cm,-1em) node[left] {\ket{1}};
    \draw[level] (2cm,-3em) -- (0cm,-3em) node[left] {\ket{0}};
    % qubit
    \draw [fill=blue] (1cm,-1.05cm) circle (.3cm);
    % pulse
    \draw[-stealth] (1cm,-1em) -- (1cm,4em);
    %% B
    % Draw the energy levels.
    \draw[level] (4.5cm,4em) -- (2.5cm,4em);
    \draw[level] (4.5cm,-1em) -- (2.5cm,-1em);
    \draw[level] (4.5cm,-3em) -- (2.5cm,-3em);
    % qubit
    \draw [fill=blue] (3.5cm,-1em) circle (.3cm);
    
    %%% 2
    \draw[->] (4.5cm,8em) -- (6cm,8em) node[midway, left, xshift=-1.1cm] {\ket{01}};
    %% A
    % Draw the energy levels.
    \draw[level] (8cm,4em) -- (6cm,4em);
    \draw[level] (8cm,-1em) -- (6cm,-1em);
    \draw[level] (8cm,-3em) -- (6cm,-3em);
    % qubit
    \draw [fill=blue] (7cm,-1.05cm) circle (.3cm);
    %% B
    % Draw the energy levels.
    \draw[level] (10.5cm,4em) -- (8.5cm,4em);
    \draw[level] (10.5cm,-1em) -- (8.5cm,-1em);
    \draw[level] (10.5cm,-3em) -- (8.5cm,-3em);
    % pulse
    \draw[-stealth] (9.5cm,-1em) -- (9.5cm,4em);
    % qubit
    \draw [fill=blue] (9.5cm,-1em) circle (.3cm);

    %%% 3
    \draw[->] (10.5cm,8em) -- (12cm,8em) node[midway, left, xshift=-1.1cm] {\ket{01}};
    %% A
    % Draw the energy levels.
    \draw[level] (14cm,4em) -- (12cm,4em);
    \draw[level] (14cm,-1em) -- (12cm,-1em);
    \draw[level] (14cm,-3em) -- (12cm,-3em);
    % qubit
    \draw [fill=blue] (13cm,-1.05cm) circle (.3cm);
    %% B
    % Draw the energy levels.
    \draw[level] (16.5cm,4em) -- (14.5cm,4em);
    \draw[level] (16.5cm,-1em) -- (14.5cm,-1em);
    \draw[level] (16.5cm,-3em) -- (14.5cm,-3em);
    % pulse
    \draw[-stealth] (15.5cm,4em) -- (15.5cm,-1em);
    % qubit
    \draw [fill=blue] (15.5cm,4em) circle (.3cm);

    %%% 4
    \draw[->] (16.5cm,8em) -- (18cm,8em) node[midway, left, xshift=-1.1cm] {i\ket{0r}};
    %% A
    % Draw the energy levels.
    \draw[level] (20cm,4em) -- (18cm,4em);
    \draw[level] (20cm,-1em) -- (18cm,-1em);
    \draw[level] (20cm,-3em) -- (18cm,-3em);
    % qubit
    \draw [fill=blue] (19cm,-1.05cm) circle (.3cm);
    %% B
    % Draw the energy levels.
    \draw[level] (22.5cm,4em) -- (20.5cm,4em);
    \draw[level] (22.5cm,-1em) -- (20.5cm,-1em);
    \draw[level] (22.5cm,-3em) -- (20.5cm,-3em);
    % pulse
    \draw[-stealth] (19cm,4em) -- (19cm,-1em);
    % qubit
    \draw [fill=blue] (21.5cm,-1em) circle (.3cm);

    %%% 5
    \draw[->] (22.5cm,8em) -- (24cm,8em) node[midway, left, xshift=-1.1cm] {-\ket{01}} node[midway, right, xshift=1.1cm] {-\ket{01}};
    %% A
    % Draw the energy levels.
    \draw[level] (26cm,4em) -- (24cm,4em);
    \draw[level] (26cm,-1em) -- (24cm,-1em);
    \draw[level] (26cm,-3em) -- (24cm,-3em);
    % qubit
    \draw [fill=blue] (25cm,-1.05cm) circle (.3cm);
    %% B
    % Draw the energy levels.
    \draw[level] (28.5cm,4em) -- (26.5cm,4em);
    \draw[level] (28.5cm,-1em) -- (26.5cm,-1em);
    \draw[level] (28.5cm,-3em) -- (26.5cm,-3em);
    % qubit
    \draw [fill=blue] (27.5cm,-1em) circle (.3cm);

    \draw[-] (-1.5cm, -5em) -- (29cm, -5em); 
    \end{tikzpicture}

    %%%% |10>
    \begin{tikzpicture}[
      scale=0.5,
      level/.style={thick},
      virtual/.style={thick,densely dashed},
      trans/.style={thick,<->,shorten >=2pt,shorten <=2pt,>=stealth},
      classical/.style={thin,double,<->,shorten >=4pt,shorten <=4pt,>=stealth}
    ]
    \tikzset{phase/.style = {draw,fill=cyan,shape=circle,minimum size=20pt,inner sep=0pt},}

    %%% 1
    %% A
    % Draw the energy levels.
    \draw[level] (2cm,4em) -- (0cm,4em) node[left] {\ket{r}} ;
    \draw[level] (2cm,-1em) -- (0cm,-1em) node[left] {\ket{1}};
    \draw[level] (2cm,-3em) -- (0cm,-3em) node[left] {\ket{0}};
    % pulse
    \draw[-stealth] (1cm,-1em) -- (1cm,4em);
    % qubit
    \draw [fill=blue] (1cm,-1em) circle (.3cm);
    %% B
    % Draw the energy levels.
    \draw[level] (4.5cm,4em) -- (2.5cm,4em);
    \draw[level] (4.5cm,-1em) -- (2.5cm,-1em);
    \draw[level] (4.5cm,-3em) -- (2.5cm,-3em);
    % qubit
    \draw [fill=blue] (3.5cm,-1.05cm) circle (.3cm);
    
    %%% 2
    \draw[->] (4.5cm,8em) -- (6cm,8em) node[midway, left, xshift=-1.1cm] {\ket{10}};
    %% A
    % Draw the energy levels.
    \draw[level] (8cm,4em) -- (6cm,4em);
    \draw[level] (8cm,-1em) -- (6cm,-1em);
    \draw[level] (8cm,-3em) -- (6cm,-3em);
    % qubit
    \draw [fill=blue] (7cm,4em) circle (.3cm);
    %% B
    % Draw the energy levels.
    \draw[level] (10.5cm,4em) -- (8.5cm,4em);
    \draw[level] (10.5cm,-1em) -- (8.5cm,-1em);
    \draw[level] (10.5cm,-3em) -- (8.5cm,-3em);
    % pulse
    \draw[-stealth] (9.5cm,-1em) -- (9.5cm,4em);
    % qubit
    \draw [fill=blue] (9.5cm,-1.05cm) circle (.3cm);

    %%% 3
    \draw[->] (10.5cm,8em) -- (12cm,8em) node[midway, left, xshift=-1.1cm] {i\ket{r0}};
    %% A
    % Draw the energy levels.
    \draw[level] (14cm,4em) -- (12cm,4em);
    \draw[level] (14cm,-1em) -- (12cm,-1em);
    \draw[level] (14cm,-3em) -- (12cm,-3em);
    % qubit
    \draw [fill=blue] (13cm,4em) circle (.3cm);
    %% B
    % Draw the energy levels.
    \draw[level] (16.5cm,4em) -- (14.5cm,4em);
    \draw[level] (16.5cm,-1em) -- (14.5cm,-1em);
    \draw[level] (16.5cm,-3em) -- (14.5cm,-3em);
    % pulse
    \draw[-stealth] (15.5cm,4em) -- (15.5cm,-1em);
    % qubit
    \draw [fill=blue] (15.5cm,-1.05cm) circle (.3cm);

    %%% 4
    \draw[->] (16.5cm,8em) -- (18cm,8em) node[midway, left, xshift=-1.1cm] {i\ket{r0}};
    %% A
    % Draw the energy levels.
    \draw[level] (20cm,4em) -- (18cm,4em);
    \draw[level] (20cm,-1em) -- (18cm,-1em);
    \draw[level] (20cm,-3em) -- (18cm,-3em);
    % pulse
    \draw[-stealth] (19cm,4em) -- (19cm,-1em);
    % qubit
    \draw [fill=blue] (19cm,4em) circle (.3cm);
    %% B
    % Draw the energy levels.
    \draw[level] (22.5cm,4em) -- (20.5cm,4em);
    \draw[level] (22.5cm,-1em) -- (20.5cm,-1em);
    \draw[level] (22.5cm,-3em) -- (20.5cm,-3em);
    % qubit
    \draw [fill=blue] (21.5cm,-1.05cm) circle (.3cm);

    %%% 5
    \draw[->] (22.5cm,8em) -- (24cm,8em) node[midway, left, xshift=-1.1cm] {i\ket{r0}} node[midway, right, xshift=1.1cm] {-\ket{10}};
    %% A
    % Draw the energy levels.
    \draw[level] (26cm,4em) -- (24cm,4em);
    \draw[level] (26cm,-1em) -- (24cm,-1em);
    \draw[level] (26cm,-3em) -- (24cm,-3em);
    % qubit
    \draw [fill=blue] (25cm,-1em) circle (.3cm);
    %% B
    % Draw the energy levels.
    \draw[level] (28.5cm,4em) -- (26.5cm,4em);
    \draw[level] (28.5cm,-1em) -- (26.5cm,-1em);
    \draw[level] (28.5cm,-3em) -- (26.5cm,-3em);
    % qubit
    \draw [fill=blue] (27.5cm,-1.05cm) circle (.3cm);

    \draw[-] (-1.5cm, -5em) -- (29cm, -5em); 
    \end{tikzpicture}

    %%%% |11>
    \begin{tikzpicture}[
      scale=0.5,
      level/.style={thick},
      virtual/.style={thick,densely dashed},
      trans/.style={thick,<->,shorten >=2pt,shorten <=2pt,>=stealth},
      classical/.style={thin,double,<->,shorten >=4pt,shorten <=4pt,>=stealth}
    ]
    \tikzset{phase/.style = {draw,fill=cyan,shape=circle,minimum size=20pt,inner sep=0pt},}

    %%% 1
    %% A
    % Draw the energy levels.
    \draw[level] (2cm,4em) -- (0cm,4em) node[left] {\ket{r}} ;
    \draw[level] (2cm,-1em) -- (0cm,-1em) node[left] {\ket{1}};
    \draw[level] (2cm,-3em) -- (0cm,-3em) node[left] {\ket{0}};
    % pulse
    \draw[-stealth] (1cm,-1em) -- (1cm,4em);
    % qubit
    \draw [fill=blue] (1cm,-1em) circle (.3cm);
    %% B
    % Draw the energy levels.
    \draw[level] (4.5cm,4em) -- (2.5cm,4em);
    \draw[level] (4.5cm,-1em) -- (2.5cm,-1em);
    \draw[level] (4.5cm,-3em) -- (2.5cm,-3em);
    % qubit
    \draw [fill=blue] (3.5cm,-1em) circle (.3cm);
    
    %%% 2
    \draw[->] (4.5cm,8em) -- (6cm,8em) node[midway, left, xshift=-1.1cm] {\ket{11}};
    %% A
    % Draw the energy levels.
    \draw[level] (8cm,4em) -- (6cm,4em);
    \draw[level] (8cm,-1em) -- (6cm,-1em);
    \draw[level] (8cm,-3em) -- (6cm,-3em);
    % qubit
    \draw [fill=blue] (7cm,4em) circle (.3cm);
    %% B
    % Draw the energy levels.
    \draw[level] (10.5cm,6em) -- (8.5cm,6em);
    \draw[level] (10.5cm,-1em) -- (8.5cm,-1em);
    \draw[level] (10.5cm,-3em) -- (8.5cm,-3em);
    \draw[virtual] (10.5cm,4em) -- (8.5cm,4em);
    % pulse
    \draw[-stealth] (9.5cm,-1em) -- (9.5cm,4em);
    % qubit
    \draw [fill=blue] (9.5cm,-1em) circle (.3cm);

    %%% 3
    \draw[->] (10.5cm,8em) -- (12cm,8em) node[midway, left, xshift=-1.1cm] {i\ket{r1}};
    %% A
    % Draw the energy levels.
    \draw[level] (14cm,4em) -- (12cm,4em);
    \draw[level] (14cm,-1em) -- (12cm,-1em);
    \draw[level] (14cm,-3em) -- (12cm,-3em);
    % qubit
    \draw [fill=blue] (13cm,4em) circle (.3cm);
    %% B
    % Draw the energy levels.
    \draw[level] (16.5cm,6em) -- (14.5cm,6em);
    \draw[level] (16.5cm,-1em) -- (14.5cm,-1em);
    \draw[level] (16.5cm,-3em) -- (14.5cm,-3em);
    \draw[virtual] (16.5cm,4em) -- (14.5cm,4em);
    % pulse
    \draw[-stealth] (15.5cm,4em) -- (15.5cm,-0.2em);
    % qubit
    \draw [fill=blue] (15.5cm,-1em) circle (.3cm);

    %%% 4
    \draw[->] (16.5cm,8em) -- (18cm,8em) node[midway, left, xshift=-1.1cm] {i\ket{r1}};
    %% A
    % Draw the energy levels.
    \draw[level] (20cm,4em) -- (18cm,4em);
    \draw[level] (20cm,-1em) -- (18cm,-1em);
    \draw[level] (20cm,-3em) -- (18cm,-3em);
    % pulse
    \draw[-stealth] (19cm,4em) -- (19cm,-1em);
    % qubit
    \draw [fill=blue] (19cm,4em) circle (.3cm);
    %% B
    % Draw the energy levels.
    \draw[level] (22.5cm,6em) -- (20.5cm,6em);
    \draw[level] (22.5cm,-1em) -- (20.5cm,-1em);
    \draw[level] (22.5cm,-3em) -- (20.5cm,-3em);
    \draw[virtual] (22.5cm,4em) -- (20.5cm,4em);
    % qubit
    \draw [fill=blue] (21.5cm,-1em) circle (.3cm);

    %%% 5
    \draw[->] (22.5cm,8em) -- (24cm,8em) node[midway, left, xshift=-1.1cm] {i\ket{r1}} node[midway, right, xshift=1.1cm] {-\ket{11}};
    %% A
    % Draw the energy levels.
    \draw[level] (26cm,4em) -- (24cm,4em);
    \draw[level] (26cm,-1em) -- (24cm,-1em);
    \draw[level] (26cm,-3em) -- (24cm,-3em);
    % qubit
    \draw [fill=blue] (25cm,-1em) circle (.3cm);
    %% B
    % Draw the energy levels.
    \draw[level] (28.5cm,4em) -- (26.5cm,4em);
    \draw[level] (28.5cm,-1em) -- (26.5cm,-1em);
    \draw[level] (28.5cm,-3em) -- (26.5cm,-3em);
    % qubit
    \draw [fill=blue] (27.5cm,-1em) circle (.3cm);
    
    \end{tikzpicture}

    \caption{Pulse sequence for a CZ gate, where each row shows the effect of the same pulse sequence on the four different basis states.}
    \label{CZpulses}
\end{figure}

As shown in figure \ref{CZpulses} with the example of the CZ gate, each atom is addressed in turn. If the atom currently addressed is in the $\ket{1}$ state, then it is excited up to the Rydberg state $\ket{r}$. However, if there is already an atom within the blockade radius in state $\ket{r}$, then the state is detuned, and the addressed atom's energy level drops back down to $\ket{1}$. The state of any atom that reaches the Rydberg level is phased flipped. After all atoms have been addressed, the atoms are then again addressed in turn to drive the states in $\ket{r}$ back down to $\ket{1}$, keeping the phase flip received during the Rydberg transition.

Current state-of-the-art experiments have achieved gate fidelities of $\mathcal{F}=0.9999$ for single qubit gates \cite{Nikolov2023}, $\mathcal{F}=0.9950$ for two-qubit gates \cite{Evered_2023,Levine_2019}, and $\mathcal{F}=0.8704$ for three qubit gates \cite{Levine_2019}. Coherent transport of entangled qubit pairs by 110 $\mu$m over a span of 200-300 $\mu$s with 100\% atom retention has been demonstrated \cite{Bluvstein_2022}.
Recently, an optical tweezer array of $>$6000 trapped neutal atom qubits with a coherence time of 12.6
seconds and room-temperature trapping lifetimes of 23 minutes was achieved \cite{manetsch_2025}.

%% file: content/method.tex
\section{Methodology}\label{sec:method}

\subsection{Gate circuits} \label{sec:QWs}

We use the quantum half-adder \cite{nielsen2002quantum} gate circuit method to algorithmically perform an encoded 1q-coin quantum walk on a $2^n$-node ring graph \cite{Douglas_2009}, as shown in Figure \ref{NL_gen}. Since we are primarily concerned with the implementation of the multiqubit shift operator as a benchmark, %with the gate requirements of the shift operator,
we define our coin operator at every step to be simply $\text{C}=\text{R}_y(\frac{\pi}{2})$ -- therefore the probabilities of the walker stepping clockwise or anti-clockwise are each $\frac{1}{2}$. The shift operator has been split into \emph{increment} and \emph{decrement}: the increment gates move the walker clockwise if applied to state $\ket{x, 1}$ and the decrement gates anticlockwise if applied to $\ket{x, 0}$. The shift operator acts on $(n+1)$ qubits, composed of two each of the set of control gates $\{ \text{C}k\text{X} |\,\,  k \in \mathbb{N} \text{ and } 1 \geq k \geq n \}$, and $2n$ X gates. Therefore, a 1q-coin QW on a $2^n$-node ring requires gates of rank $1 \geq r \geq n+1$, where the rank $r$ of a gate is the number of qubits it acts on. 

% NL general
\begin{figure}[h!]
    \centering
    \begin{tikzpicture}[thick]
        \tikzset{
        operator/.style = {draw,fill=white,minimum size=0.1em}, 
        dots/.style = {draw=white,fill=white,minimum height=0.5cm, minimum width=1cm},
        operator2/.style = {draw,fill=white,minimum height=2cm, minimum width=1cm}, 
        phase/.style = {draw,fill,shape=circle,minimum size=5pt,inner sep=0pt}, 
        surround/.style = {fill=blue!10,thick,draw=black,rounded corners=2mm}, 
        cross/.style={path picture={\draw[thick,black](path picture bounding box.north) -- (path picture bounding box.south) (path picture bounding box.west) -- (path picture bounding box.east); }}, crossx/.style={path picture={ \draw[thick,black,inner sep=0pt] (path picture bounding box.south east) -- (path picture bounding box.north west) (path picture bounding box.south west) -- (path picture bounding box.north east); }}, 
        circlewc/.style={draw,circle,cross,minimum width=0.3 cm}, 
        meter/.style= {draw, fill=white, inner sep=5, rectangle, font=\vphantom{A}, minimum width=20, line width=.8, path picture={\draw[black] ([shift={(.1,.2)}]path picture bounding box.south west) to[bend left=30] ([shift={(-.1,.2)}]path picture bounding box.south east);\draw[black,-latex] ([shift={(0,.1)}]path picture bounding box.south) -- ([shift={(.2,-.1)}]path picture bounding box.north);}}, }
        \matrix[row sep=0.2cm, column sep=0.35cm] (circuit) {&\node(lt)[] {coin}; &\node (l1up) {}; & &[-6mm] \node(lt) {increment}; &[-6mm] && \node (l2up) {}; & && &[-6mm] \node(lt) {decrement}; &[-3mm]; % LINE 0up
        \\ 
        \node (q1) {$x_1$: $\ket{0}$}; % LINE x1
        &&& \node[circlewc] (C1) {}; &\node[dots] (d) {\dots}; &&&&\node[operator] (H0) {X}; &\node[operator] (H0) {X}; &\node[circlewc] (C2) {}; &\node[dots] (m) {$\dots$};&&& &\coordinate (end1); &[-0.2cm]\node (d1) {$\cdots$}; &[-0.3cm]\node[meter] (meter) {};
        \\ 
        \node (qm) {$\vdots$}; % LINE xm
        &&&\node[dots] (m) {$\vdots$}; &\node (m) {$\ddots$}; &&&&\node[dots] (m) {$\vdots$};&&\node[dots] (m) {$\vdots$}; &\node (m) {$\ddots$}; &&&& \coordinate (endm); &[-0.2cm]\node (dm) {$\cdots$}; &[-0.3cm]\node[meter] (meter) {};
        \\
        \node (q2) {$x_n$: $\ket{0}$}; % LINE xn
        &&&\node[phase] (P0) {}; &\node[dots] (d) {\dots}; && \node[circlewc] (C21) {}; &&\node[operator] (H0) {X}; &&\node[phase] (P0) {}; &\node[dots] (m) {$\dots$}; &\node[operator] (H0) {X}; &\node[circlewc] (C22) {}; && \coordinate (end2); &[-0.2cm]\node (d1) {$\cdots$}; &[-0.3cm]\node[meter] (meter) {};
        \\ \node (q3) {$c_1$: $\ket{0}$}; % LINE c1
        &\node[operator] (H11) {$\text{R}_y(\frac{\pi}{2})$}; &&\node[phase] (P2) {}; &\node[dots] (m) {$\dots$}; &&\node[phase] (P1) {}; &&\node[operator] (X1) {X}; &&\node[phase] (P11) {}; &\node[dots] (m) {$\dots$}; &&\node[phase] (P3) {}; &\node[operator] (H0) {X}; & \coordinate (end3); &[-0.2cm]\node (d1) {$\cdots$}; \\
        && \node (l1dn) {}; &&&&& \node (l2dn) {};  % LINE 0dn
        \\ 
        };
        \begin{pgfonlayer}{background} 
        \draw[thick] (q1) -- (end1) (q2) -- (end2) (q3) -- (end3) (P1) -- (C21) (P2) -- (C1) (P3) -- (C22) (P11) -- (C2);
        \draw[thick, dotted] (l1up) -- (l1dn) (l2up) -- (l2dn);
        \end{pgfonlayer}
    \end{tikzpicture}
    \caption{Gate circuit for one-qubit coin (1q-coin) quantum walk on a $2^n$-node ring \cite{Douglas_2009}. The multiqubit gates are C$k$X gates, where the small circles denote control qubits and the circled crosses denote target qubits.}
    \label{NL_gen}
\end{figure}

%For the eventual application of quantum walks to fluid dynamics the walker also requires a rest state -- the possibility of no movement -- which [enacts] a `lazy' quantum walk. In order to [enact] a lazy quantum walk, the non-lazy QW circuit can be adapted by increasing the dimension of the coin. For example, Ref. \cite{Saha_2021} presents a lazy QW using a ternary coin and shift operator s.t. $\text{S} \ket{x, 0} = \ket{x, 0} \text{, S} \ket{x, 1} = \ket{x-1, 1}\text{, and S} \ket{x, 2} = \ket{x+1, 2}$. This work uses qubits exclusively, and so the coin and shift operators must be modified accordingly (see Section \ref{lazy_circuits}).

Figure \ref{L_gen} shows the circuit for a 2q-coin quantum walk on a $2^n$-node ring graph -- the gates from the respective 1q-coin circuits are controlled on both coin qubits, with the exception of the one-qubit X gates as the controls cancel out. Therefore, a 2q-coin QW on a $2^n$-node ring requires gates of rank one and of rank $3 \leq r \leq n+2$. The coin operator is simply $\text{R}_y(\frac{\pi}{2})$ applied to each coin qubit, therefore at each step the walker will rest with probability $\frac{1}{2}$, move clockwise with probability $\frac{1}{4}$, and move anticlockwise with probability $\frac{1}{4}$.
%In general, an $n_c$-qubit coin QW on a $2^n$-node ring requires gates of rank one and of rank $1+n_c \leq r \leq n_q$ where $n_q=n+n_c$.

% L general
\begin{figure}[h!]
    \centering
    \begin{tikzpicture}[thick]
        \tikzset{
        operator/.style = {draw,fill=white,minimum size=0.1em}, 
        dots/.style = {draw=white,fill=white,minimum height=0.5cm, minimum width=1cm},
        operator2/.style = {draw,fill=white,minimum height=2cm, minimum width=1cm}, 
        phase/.style = {draw,fill,shape=circle,minimum size=5pt,inner sep=0pt}, 
        surround/.style = {fill=blue!10,thick,draw=black,rounded corners=2mm}, 
        cross/.style={path picture={\draw[thick,black](path picture bounding box.north) -- (path picture bounding box.south) (path picture bounding box.west) -- (path picture bounding box.east); }}, crossx/.style={path picture={ \draw[thick,black,inner sep=0pt] (path picture bounding box.south east) -- (path picture bounding box.north west) (path picture bounding box.south west) -- (path picture bounding box.north east); }}, 
        circlewc/.style={draw,circle,cross,minimum width=0.3 cm}, 
        meter/.style= {draw, fill=white, inner sep=5, rectangle, font=\vphantom{A}, minimum width=20, line width=.8, path picture={\draw[black] ([shift={(.1,.2)}]path picture bounding box.south west) to[bend left=30] ([shift={(-.1,.2)}]path picture bounding box.south east);\draw[black,-latex] ([shift={(0,.1)}]path picture bounding box.south) -- ([shift={(.2,-.1)}]path picture bounding box.north);}}, }
        \matrix[row sep=0.2cm, column sep=0.35cm] (circuit) {&\node(lt)[] {coin}; &\node (l1up) {}; & &[-6mm] \node(lt) {increment}; &[-6mm] && \node (l2up) {}; & && &[-6mm] \node(lt) {decrement}; &[-3mm]; % LINE 0up
        \\ 
        \node (q1) {$x_1$: $\ket{0}$}; % LINE x1
        &&& \node[circlewc] (C1) {}; &\node[dots] (d) {\dots}; &&&&\node[operator] (H0) {X}; &\node[operator] (H0) {X}; &\node[circlewc] (C2) {}; &\node[dots] (m) {$\dots$};&&& &\coordinate (end1); &[-0.2cm]\node (d1) {$\cdots$}; &[-0.3cm]\node[meter] (meter) {};
        \\ 
        \node (qm) {$\vdots$}; % LINE xm
        &&&\node[dots] (m) {$\vdots$}; &\node (m) {$\ddots$}; &&&&\node[dots] (m) {$\vdots$};&&\node[dots] (m) {$\vdots$}; &\node (m) {$\ddots$}; &&&& \coordinate (endm); &[-0.2cm]\node (dm) {$\cdots$}; &[-0.3cm]\node[meter] (meter) {};
        \\
        \node (q2) {$x_n$: $\ket{0}$}; % LINE xn
        &&&\node[phase] (P0) {}; &\node[dots] (d) {\dots}; && \node[circlewc] (C21) {}; &&\node[operator] (H0) {X}; &&\node[phase] (P0) {}; &\node[dots] (m) {$\dots$}; &\node[operator] (H0) {X}; &\node[circlewc] (C22) {}; && \coordinate (end2); &[-0.2cm]\node (d1) {$\cdots$}; &[-0.3cm]\node[meter] (meter) {};
        \\ \node (startc) {$c_1$: $\ket{0}$}; % LINE c1
        &\node[operator] (H11) {$\text{R}_y(\frac{\pi}{2})$}; &&\node[phase] (P2) {}; &\node[dots] (m) {$\dots$}; &&\node[phase] (P1) {}; &&\node[operator] (X1) {X}; &&\node[phase] (P11) {}; &\node[dots] (m) {$\dots$}; &&\node[phase] (P3) {}; &\node[operator] (H0) {X}; & \coordinate (endc); &[-0.2cm]\node (d1) {$\cdots$}; \\
        && \node (l1dn) {}; &&&&& \node (l2dn) {};  % LINE 0dn
        \\ 
        \node (q3) {$c_2$: $\ket{0}$}; % LINE c1
        &\node[operator] (H11) {$\text{R}_y(\frac{\pi}{2})$}; &&\node[phase] (P2) {}; &\node[dots] (m) {$\dots$}; &&\node[phase] (P1) {}; && &&\node[phase] (P11) {}; &\node[dots] (m) {$\dots$}; &&\node[phase] (P3) {}; & & \coordinate (end3); &[-0.2cm]\node (d1) {$\cdots$}; \\
        && \node (l1dn) {}; &&&&& \node (l2dn) {};  % LINE 0dn
        \\ 
        };
        \begin{pgfonlayer}{background} 
        \draw[thick] (startc) -- (endc) (q1) -- (end1) (q2) -- (end2) (q3) -- (end3) (P1) -- (C21) (P2) -- (C1) (P3) -- (C22) (P11) -- (C2);
        \draw[thick, dotted] (l1up) -- (l1dn) (l2up) -- (l2dn);
        \end{pgfonlayer}
    \end{tikzpicture}
    \caption{Gate circuit for a two-qubit coin (2q-coin) quantum walk on a $2^n$-node ring. The multiqubit gates are C$k$X gates, where the small circles denote control qubits and the circled crosses denote target qubits.}
    \label{L_gen}
\end{figure}

We simulate both 1q-coin and 2q-coin quantum walks with our realistic error model, as both are good benchmarks, but bearing in mind that the 2q-coin quantum walk has greater applicability in fluid simulation.
%both to compare the results between the two and because we reason the 1q-coin will be easier to realise in the near-term.

\subsection{Error model} \label{neutral-atoms}

In this work, we consider a programmable 2D array of cesium (Cs) atoms trapped in a triangular or rectangular array separated by $d=4$ $\mu$m at temperature $T=0$ K. Sources of error considered are imperfect gate operations, state preparation, readout, and passive noise.

We consider near-term ARP (adiabatic rapid passage) Rydberg gate modelling based on the work of \citeauthor{Pelegri_2022} \cite{Pelegri_2022}. Considering an array of Cs atoms and current technologies, the authors use a detailed model that accounts for the effects of spontaneous decay and AC Stark shifts for experimentally feasible parameters. The authors therefore obtain realistic parameters for native two-qubit and multiqubit gates, with effective matrices as follows \cite{Pelegri_2022}:
\begin{align}
    \text{CZ}_\text{eff} =& \ketbra{00}{00} + 0.9990 e^{0.9906 i \pi} \left(\ketbra{01}{01} + \ketbra{10}{10}\right) + 0.9986 e^{i \pi} \ketbra{11}{11} \label{czeff} \\
    \text{CCZ}_\text{eff} =& \ketbra{000}{000} + 0.9981 e^{0.9845 i \pi} \left(\ketbra{001}{001} + \ketbra{010}{010} + \ketbra{100}{100} \right) \nonumber \label{cczeff} \\
    &+ 0.9973 e^{0.9934 i \pi}\left(\ketbra{011}{011} + \ketbra{101}{101} + \ketbra{110}{110} \right) + 0.9963 e^{0.9911 i \pi} \ketbra{111}{111} .
\end{align}
These effective matrices are completely-positive and trace-nonincreasing (CPTN) map operators with $\text{CkZ}_\text{eff}^\dagger\text{CkZ}_\text{eff}\leq I$, and therefore population leakage is modelled by loss of total probability associated with the qubits' state vector. (We therefore do not renormalise at any point in the computation to preserve this probability loss.)

Although not yet experimentally implemented, the authors also present the theoretical effective matrix for the C3Z gate \cite{Pelegri_2022}:
\begin{align} \label{c3zeff}
    \text{C3Z}_\text{eff} = \ketbra{0000}{0000} &+ 0.997947 e^{-0.995 i \pi} \sum_{\sigma \in \mathbf{H}^4_1} \ketbra{\sigma}{\sigma}  
    + 0.996286 e^{0.984 i \pi} \sum_{\sigma \in \mathbf{H}^4_2} \ketbra{\sigma}{\sigma} \nonumber \\
    &+ 0.994391 e^{0.981 i \pi} \sum_{\sigma \in \mathbf{H}^4_3} \ketbra{\sigma}{\sigma} 
    + 0.990724 e^{0.981 i \pi} \ketbra{1111}{1111}
\end{align}
where $\mathbf{H}^4_a$ is the set of all bit strings of length 4 with Hamming weight (population count of state 1) $a$. Therefore the gate fidelity of each is $\mathcal{F}(\text{CZ}_\text{eff}) = 0.9981$, $\mathcal{F}(\text{CCZ}_\text{eff}) = 0.9954$, $\mathcal{F}(\text{C3Z}_\text{eff}) = 0.9850$. %(In a native two-qubit gate setting these multiqubit gate fidelities could be recovered \cite{Barenco_1995} with two-qubit gate fidelities of $\mathcal{F}(\text{CZ}) = \sqrt[5]{\mathcal{F}(\text{CCZ}_\text{eff})} = 0.9991$ and $\mathcal{F}(\text{CZ}) = \sqrt[20]{\mathcal{F}(\text{C3Z}_\text{eff})} = 0.9992$ -- see Section \ref{sec:decomp}.)

%(Compare this to competitive two-qubit gates in superconducting hardware with fidelity $\mathcal{F}(\text{CZ})=0.9990$ [CITE], which clearly exceeds the neutral atom two-qubit gate fidelity. However, in a superconducting scheme, a three-qubit gate must be decomposed into five \cite{Barenco_1995} two-qubit gates with fidelity $\mathcal{F}(\text{CCZ})\leq 0.9990^5 =0.9950$ $0.9993^5=0.9965$ and equivalently a four-qubit gate into twenty two-qubit gates such that $\mathcal{F}(\text{C3Z})\leq 0.9990^{20} =0.9800$ $0.9993^20=0.9861$. Therefore the fidelity of three-qubit and larger gates is greater for neutral atom hardware than for platforms without native multi-qubit gates. Furthermore, multi-qubit gates drastically reduce gate depth. Note that if one is implementing a quantum walk on a device restricted to two-qubit gates, it would be beneficial to chose a different gate circuit from the ones described in this work, such as the circuit in Ref \cite{Shakeel_2020} which uses only 2-qubit controlled rotations. However a direct comparison with this scheme is outside of the scope of this work.)

A $\text{C}k\text{Z}$ gate can be transformed into the required $\text{C}k\text{X}$ gate with the addition of single-qubit gates, where:
\begin{align}\label{CZtoCX}
    \text{C}k \text{X} = (\text{R}^{2 \pi}_z)^{\otimes k+1} (\text{Z}^{\otimes k}\otimes \text{ZHZ}) \text{C}k \text{Z} (\text{R}^{2 \pi}_z)^{\otimes k+1} (\text{Z}^{\otimes k}\otimes \text{ZHZ})
\end{align}
where the target is the last qubit, $\text{R}^{2 \pi}_z \equiv -\text{I}_2$, and $\text{H} = \frac{1}{\sqrt{2}} \begin{pmatrix} 1 & 1 \\ 1 & -1 \end{pmatrix}$. Since single-qubit gates have negligible error compared to multiqubit gates ($\mathcal{F}(\text{H,Z,X})=0.9999$), we model single-qubit gates with their ideal gates, and therefore $\mathcal{F}(\text{C}k\text{X}_\text{eff})=\mathcal{F}(\text{C}k\text{Z}_\text{eff})$.

Further errors considered in this work are state preparation, readout, waiting times, and qubit movement. We use the depolarising channel, dephasing channel, CPTN maps, and random number generators to model longitudinal relaxation, transversal relaxation, population leakage, and atom loss. 
Due to the use of weighted random number generators, we repeat each QW 1000 times (if $n_q \geq 5$) or 2000 times (if $n_q \leq 4$) with new random numbers for each qubit and error source, and then take the mean of the outputted state fidelities.

The depolarising channel \cite{Gustiani_2025} is the density matrix mapping
\begin{align}
    \rho \rightarrow (1-\epsilon) \rho + \frac{\epsilon}{3} \sum_\sigma \sigma \rho \sigma
\end{align}
where $\sigma \in \{$X, Y, Z$\}$ are the Pauli operators. We approximate this channel by applying one of the operators in the set $\text{E}_1(\epsilon) \in \{$I,X,Y,Z$\}$ according to a random number generator, weighted such that $\text{Prob}(\text{E}_1(\epsilon) = \text{I}) = 1 - \epsilon$ and $\text{Prob}(\text{E}_1(\epsilon) = \text{X}) = \text{Prob}(\text{E}_1(\epsilon) = \text{Y}) = \text{Prob}(\text{E}_1(\epsilon) = \text{Z}) =  \frac{\epsilon}{3}$.

Similarly, the dephasing channel \cite{Gustiani_2025} with mapping
\begin{align}
    \rho \rightarrow (1-\epsilon) \rho + \epsilon Z \rho Z
\end{align}
is approximated by applying $\text{E}_2(\epsilon) \in \{$I,Z$\}$ such that $\text{Prob}(\text{E}_2(\epsilon) = \text{I}) = 1 - \epsilon$ and $\text{Prob}(\text{E}_2(\epsilon) = \text{Z}) =  \epsilon$.

Passive noise -- the effect of decay processes experienced by qubits while they wait for other qubits to be acted on -- is characterised by longitudinal $T_1$ and transversal $T_2$ relaxation times. $T_1$ decay is modelled using depolarising noise and therefore applied with $\text{E}_1(\epsilon_\text{wait1})$ where \cite{Gustiani_2025}
\begin{align}
    \epsilon_\text{wait1}(\Delta t) = \frac{3}{4} \left( 1 - \exp \left[ - \frac{\Delta t}{T_1} \right] \right) .
\end{align}
for duration $\Delta t$ where $T_1=4s$ \cite{Bluvstein_2022}. The $T_2$ decay is modelled with dephasing noise by applying $\text{E}_2(\epsilon_\text{wait2})$ where \cite{Gustiani_2025}
\begin{align}
    \epsilon_\text{wait2}(\Delta t) = \frac{1}{2} \left( 1 - \exp \left[ - \frac{\Delta t}{T_2} \right] \right) .
\end{align}
for duration $\Delta t$ where $T_2=1.49s$ \cite{Bluvstein_2022}. Therefore $\text{E}_\text{wait}(\Delta t) = \text{E}_1(\epsilon_\text{wait1}(\Delta t))\text{E}_2(\epsilon_\text{wait2}(\Delta t))$ is applied to each qubit not otherwise being acted on for duration $\Delta t$.

At the start of each QW, each qubit is initialised in state $\ket{0}$ but subject to population leakage with probability $\epsilon_\text{init}=0.003$ due to finite optical pumping \cite{Evered_2023}. This leakage is modelled \cite{Gustiani_2025} simply by applying the CPTN map operator
\begin{align}
    \text{E}_\text{init}(\epsilon_\text{init}) = \begin{pmatrix} \sqrt{1-\epsilon_\text{init}} & 0 \\ 0 & 1 \end{pmatrix}
\end{align}
to each qubit.

During the walk, qubits may need to be moved to a different space in the array in order to interact with the required qubits. If this movement is performed slowly enough -- $v \leq 0.55 \,\mu$m$(\mu\text{s})^{-1}$ -- then no fidelity penalty is incurred \cite{Bluvstein_2022}. Therefore (below this speed) the only affect of movement is the passive noise. For qubits waiting during a movement, we choose $\Delta t = \tau_\text{move} = 100 \,\mu$s, which allows for $55 \,\mu$m of errorless movement, which is far above sufficient for all mapped out moves on a $\sim 5 \,\mu$m trap array. (See appendices \ref{app:circ3} and \ref{app:circ4} for gate circuits for each size QW, where single qubit gate `M' denotes a single atom movement.) For simplicity, every qubit undergoes passive noise applied with $\text{E}_\text{wait}(\tau_\text{move})$ during a movement. For qubits waiting \cite{Pelegri_2022} during a gate, they are subject to passive noise for gate duration $\Delta t = \tau_\text{gate} = 1.8 \,\mu$s, applied with $\text{E}_\text{wait}(\tau_\text{gate})$.

Imperfect readout is modelled with depolarising error and random atom loss. For a simulation of $t$ steps, each qubit is subject to depolarising probability $\epsilon_\text{read}=0.0017$ at the end \cite{Bluvstein_2022} applied with $\text{E}_1(\epsilon_\text{read})$. We model random atom loss, with a probability \cite{Kwon_2017} of $\epsilon_\text{loss}=0.013$, with a (weighted) random number generator for each position qubit. If an atom is lost, it is traced out of the composite state and replaced by a maximally mixed state.

\subsection{Gate decompositions} \label{sec:decomp}

We consider circuits with effective $(r\leq6)$-qubit gates, but are only considering the implementation of native gates of rank $r\leq3$ for near-term and $r\leq4$ for further term implementation on neutral atom hardware -- therefore many considered circuits must be decomposed into equivalent gate sequences composed of gates of a lesser rank. Examples of gate decompositions \cite{Barenco_1995} are shown in Figure \ref{C4Xdecomp}. A decomposition of a C$k$X gate into CCX gates requires $4(k-2)$ CCX gates and $k-2$ ancilla qubits. Decomposing a C4X(C5X) gate into native gates of rank $r\leq4$ requires two(zero) CCX gates, two(four) C3X gates, and one ancilla qubit.

\begin{figure}[h!]
    \centering
    \begin{subfigure}{0.56\textwidth}
        \begin{tikzpicture}[thick]
            \tikzset{
            operator/.style = {draw,fill=white,minimum size=0.05em}, 
            operator2/.style = {draw,fill=white,minimum height=2cm, minimum width=1cm}, 
            phase/.style = {draw,fill,shape=circle,minimum size=5pt,inner sep=0pt}, 
            surround/.style = {fill=blue!10,thick,draw=black,rounded corners=2mm}, 
            cross/.style={path picture={\draw[thick,black](path picture bounding box.north) -- (path picture bounding box.south) (path picture bounding box.west) -- (path picture bounding box.east); }}, crossx/.style={path picture={ \draw[thick,black,inner sep=0pt] (path picture bounding box.south east) -- (path picture bounding box.north west) (path picture bounding box.south west) -- (path picture bounding box.north east); }}, 
            circlewc/.style={draw,circle,cross,minimum width=0.3 cm}, 
            meter/.style= {draw, fill=white, inner sep=5, rectangle, font=\vphantom{A}, minimum width=20, line width=.8, path picture={\draw[black] ([shift={(.1,.2)}]path picture bounding box.south west) to[bend left=30] ([shift={(-.1,.2)}]path picture bounding box.south east);\draw[black,-latex] ([shift={(0,.1)}]path picture bounding box.south) -- ([shift={(.2,-.1)}]path picture bounding box.north);}}, }
            \matrix[row sep=0.2cm, column sep=0.25cm] (circuit) {
            \\\node (q1) {$q_1$:}; % LINE x1
            &\node[circlewc] (C11) {}; &\coordinate (end1);
            \\ \node (q2) {$q_2$:}; % LINE x2
            &\node[phase] (P0) {}; &\coordinate (end2);
            \\ \node (q3) {$q_3$:}; % LINE x3
            &\node[phase] (P0) {}; &\coordinate (end3);
            \\ \node (q4) {$q_4$:}; % LINE c1
            &\node[phase] (P0) {}; &\coordinate (end4); &\node (e) {$=$};
            \\ \node (q5) {$q_5$:}; % LINE c2
            &\node[phase] (P11) {}; &\coordinate (end5);
            \\
            \node (q6) {$a_1$:}; % LINE a1
            & &\coordinate (end6);
            \\ \node (q7) {$a_2$:}; % LINE a2
            & &\coordinate (end7);
            \\};
            \begin{pgfonlayer}{background} 
            \draw[thick] (q1) -- (end1) (q2) -- (end2) (q3) -- (end3) (q4) -- (end4) (q5) -- (end5) (q6) -- (end6) (q7) -- (end7)
            (P11) -- (C11);
            ; \end{pgfonlayer}
        \end{tikzpicture}
        \hspace{-0.4cm}
        \begin{tikzpicture}[thick]
            \tikzset{
            operator/.style = {draw,fill=white,minimum size=0.05em}, 
            operator2/.style = {draw,fill=white,minimum height=2cm, minimum width=1cm}, 
            phase/.style = {draw,fill,shape=circle,minimum size=5pt,inner sep=0pt}, 
            surround/.style = {fill=blue!10,thick,draw=black,rounded corners=2mm}, 
            cross/.style={path picture={\draw[thick,black](path picture bounding box.north) -- (path picture bounding box.south) (path picture bounding box.west) -- (path picture bounding box.east); }}, crossx/.style={path picture={ \draw[thick,black,inner sep=0pt] (path picture bounding box.south east) -- (path picture bounding box.north west) (path picture bounding box.south west) -- (path picture bounding box.north east); }}, 
            circlewc/.style={draw,circle,cross,minimum width=0.3 cm}, 
            meter/.style= {draw, fill=white, inner sep=5, rectangle, font=\vphantom{A}, minimum width=20, line width=.8, path picture={\draw[black] ([shift={(.1,.2)}]path picture bounding box.south west) to[bend left=30] ([shift={(-.1,.2)}]path picture bounding box.south east);\draw[black,-latex] ([shift={(0,.1)}]path picture bounding box.south) -- ([shift={(.2,-.1)}]path picture bounding box.north);}}, }
            \matrix[row sep=0.2cm, column sep=0.35cm] (circuit) {
            \\\node (q1) {$q_1$:}; % LINE x1
            &\node[circlewc] (C11) {}; &&& &\node[circlewc] (C12) {}; &&& &\coordinate (end1);
            \\ \node (q2) {$q_2$:}; % LINE x2
            &\node[phase] (P21) {}; &&& &\node[phase] (P22) {}; &&&
            &\coordinate (end2);
            \\ \node (q3) {$q_3$:}; % LINE x3
            & &\node[phase] (P31) {}; & &\node[phase] (P32) {}; & &\node[phase] (P33) {}; & &\node[phase] (P34) {};
            &\coordinate (end3);
            \\ \node (q4) {$q_4$:}; % LINE c1
            && &\node[phase] (Pc11) {}; &&&
            &\node[phase] (Pc12) {}; &
            &\coordinate (end4);
            \\ \node (q5) {$q_5$:}; % LINE c2
            && &\node[phase] (Pc21) {}; &&&
            &\node[phase] (Pc22) {}; &
            &\coordinate (end5);
            \\
            \node (q6) {$a_1$:}; % LINE a1
            & &\node[phase] (Pa11) {}; &\node[circlewc] (Ca11) {};  &\node[phase] (Pa12) {}; &
            &\node[phase] (Pa13) {}; &\node[circlewc] (Ca12) {}; &\node[phase] (Pa14) {};
            &\coordinate (end6);
            \\ \node (q7) {$a_2$:}; % LINE a2
            &\node[phase] (Pa21) {}; &\node[circlewc] (Ca21) {}; &
            &\node[circlewc] (Ca22) {};
            &\node[phase] (Pa22) {}; 
            &\node[circlewc] (Ca23) {}; &
            &\node[circlewc] (Ca24) {};
            &\coordinate (end7);
            \\};
            \begin{pgfonlayer}{background} 
            \draw[thick] (q1) -- (end1) (q2) -- (end2) (q3) -- (end3) (q4) -- (end4) (q5) -- (end5) (q6) -- (end6) (q7) -- (end7)
            (Pc11) -- (Ca11)
            (P31) -- (Ca21) (Pa21) -- (C11)
            (P32) -- (Ca22) (Pa22) -- (C12)
            (Pc12) -- (Ca12)
            (P33) -- (Ca23)
            (P34) -- (Ca24);
            ; \end{pgfonlayer}
        \end{tikzpicture}
        \caption{Effective C4X gate decomposed into native CCX gates.}
    \end{subfigure}
    \begin{subfigure}{0.43\textwidth}
        \begin{tikzpicture}[thick]
            \tikzset{
            operator/.style = {draw,fill=white,minimum size=0.1em}, 
            operator2/.style = {draw,fill=white,minimum height=2cm, minimum width=1cm}, 
            phase/.style = {draw,fill,shape=circle,minimum size=5pt,inner sep=0pt}, 
            surround/.style = {fill=blue!10,thick,draw=black,rounded corners=2mm}, 
            cross/.style={path picture={\draw[thick,black](path picture bounding box.north) -- (path picture bounding box.south) (path picture bounding box.west) -- (path picture bounding box.east); }}, crossx/.style={path picture={ \draw[thick,black,inner sep=0pt] (path picture bounding box.south east) -- (path picture bounding box.north west) (path picture bounding box.south west) -- (path picture bounding box.north east); }}, 
            circlewc/.style={draw,circle,cross,minimum width=0.3 cm}, 
            meter/.style= {draw, fill=white, inner sep=5, rectangle, font=\vphantom{A}, minimum width=20, line width=.8, path picture={\draw[black] ([shift={(.1,.2)}]path picture bounding box.south west) to[bend left=30] ([shift={(-.1,.2)}]path picture bounding box.south east);\draw[black,-latex] ([shift={(0,.1)}]path picture bounding box.south) -- ([shift={(.2,-.1)}]path picture bounding box.north);}}, }
            \matrix[row sep=0.2cm, column sep=0.25cm] (circuit) {
            \node (q1) {$q_1$:}; % LINE x1
            &\node[circlewc] (Ctop1) {}; &\coordinate (end1);
            \\ \node (q2) {$q_2$:}; % LINE x2
            &\node[phase] (P0) {}; &\coordinate (end2);
            \\ \node (q3) {$q_3$:}; % LINE x3
            &\node[phase] (P0) {}; &\coordinate (end3); &\node (e) {$=$};
            \\ \node (q4) {$q_4$:}; % LINE x4
            &\node[phase] (Pc1) {}; &\coordinate (end4);
            \\ \node (q5) {$q_5$:}; % LINE c1
            &\node[phase] (Pa1) {}; &\coordinate (end5);
            \\ \node (q6) {$a_1$:}; % LINE a1
            & &\coordinate (end6);
            \\ \\};
            \begin{pgfonlayer}{background} 
            \draw[thick] (q1) -- (end1) (q2) -- (end2) (q3) -- (end3) (q4) -- (end4) (q5) -- (end5) (q6) -- (end6)
            (Ctop1) -- (Pa1)
            ; \end{pgfonlayer}
        \end{tikzpicture}
        \hspace{-0.4cm}
        \begin{tikzpicture}[thick]
            \tikzset{
            operator/.style = {draw,fill=white,minimum size=0.1em}, 
            operator2/.style = {draw,fill=white,minimum height=2cm, minimum width=1cm}, 
            phase/.style = {draw,fill,shape=circle,minimum size=5pt,inner sep=0pt}, 
            surround/.style = {fill=blue!10,thick,draw=black,rounded corners=2mm}, 
            cross/.style={path picture={\draw[thick,black](path picture bounding box.north) -- (path picture bounding box.south) (path picture bounding box.west) -- (path picture bounding box.east); }}, crossx/.style={path picture={ \draw[thick,black,inner sep=0pt] (path picture bounding box.south east) -- (path picture bounding box.north west) (path picture bounding box.south west) -- (path picture bounding box.north east); }}, 
            circlewc/.style={draw,circle,cross,minimum width=0.3 cm}, 
            meter/.style= {draw, fill=white, inner sep=5, rectangle, font=\vphantom{A}, minimum width=20, line width=.8, path picture={\draw[black] ([shift={(.1,.2)}]path picture bounding box.south west) to[bend left=30] ([shift={(-.1,.2)}]path picture bounding box.south east);\draw[black,-latex] ([shift={(0,.1)}]path picture bounding box.south) -- ([shift={(.2,-.1)}]path picture bounding box.north);}}, }
            \matrix[row sep=0.2cm, column sep=0.35cm] (circuit) {
            \node (q1) {$q_1$:}; % LINE x1
            & &\node[circlewc] (Ctop1) {}; &&\node[circlewc] (Ctop2) {}; &\coordinate (end1);
            \\ \node (q2) {$q_2$:}; % LINE x2
            &&\node[phase] (P0) {}; &&\node[phase] (P0) {}; &\coordinate (end2);
            \\ \node (q3) {$q_3$:}; % LINE x3
             &&\node[phase] (P0) {}; &&\node[phase] (P0) {}; &\coordinate (end3);
            \\ \node (q4) {$q_4$:}; % LINE x4
            &\node[phase] (Pc1) {}; &&\node[phase] (Pc2) {}; &&\coordinate (end4);
            \\ \node (q5) {$q_5$:}; % LINE c1
            &\node[phase] (P1) {}; &&\node[phase] (P1) {}; &&\coordinate (end5);
            \\ \node (q6) {$a_1$:}; % LINE a1
            &\node[circlewc] (Ca1) {}; &\node[phase] (Pa1) {};
            &\node[circlewc] (Ca2) {}; &\node[phase] (Pa2) {};
            &\coordinate (end6);
            \\ \\};
            \begin{pgfonlayer}{background} 
            \draw[thick] (q1) -- (end1) (q2) -- (end2) (q3) -- (end3) (q4) -- (end4) (q5) -- (end5) (q6) -- (end6)
            (Ctop1) -- (Pa1)
            (Ctop2) -- (Pa2)
            (Ca1) -- (Pc1)
            (Ca2) -- (Pc2)
            ; \end{pgfonlayer}
        \end{tikzpicture}
        \caption{Effective C4X gates decomposed into native C3X and CCX gates.}
    \end{subfigure}
    \caption{Gate decomposition examples \cite{Barenco_1995} for C4X. Qubits $q_j$ are qubits undergoing the gate operation and qubits $a_j$ are ancilla. %Qubits $\{q_1 ... q_5\}$ are the qubits undergoing the operation C$k$X, with $q_1$ as the target, and qubits $\{a_1, a_2\}$ are ancilla.
    }
    \label{C4Xdecomp}
\end{figure}

The decomposition of each circuit used for the error modelling in this work are shown in Appendix \ref{app:circ3} for native gates of maximum rank 3 and Appendix \ref{app:circ4} for native gates of maximum rank 4. Atoms are considered to be on a 2D array with maximum static connectivity of four atoms each, and any atom movement required is denoted with the single qubit gate `M'.

Although not used in this work, multiqubit C$k$X(C$k$Z) gates can also be decomposed into CX(CZ) gates. For example, a CCX gate can be implemented with five native CX gates \cite{Barenco_1995}, resulting in $\mathcal{F}(\text{CCX})_\text{eff} \approx \mathcal{F}(\text{CX}_\text{eff})^5$. With this logic we can backwards engineer the native two-qubit fidelities required to recover the native multiqubit fidelities used in our model, specifically: $\mathcal{F}(\text{CX}_\text{eff}) = \sqrt[5]{\mathcal{F}(\text{CCX}_\text{eff})} = 0.9991$ and $\mathcal{F}(\text{CX}_\text{eff}) = \sqrt[20]{\mathcal{F}(\text{C3X}_\text{eff})} = 0.9992$, which aligns with state-of-the-art 40 hr-averaged \cite{marxer2025999fidelitysinglequbitgates} and exceeds state-of-the-art fast \cite{Fasciati-2025} CZ gate implementations in superconducting hardware. (Of course, detailed realistic error modelling with SPAM, movement, and CPTN maps is required for accurate comparison.)

%% file: content/results.tex
\section{Results}\label{sec:results}
\subsection{Near-term}\label{near-term}

We perform a numerical simulation (on a laptop with an Intel core i5 processor) of 21 steps of one-qubit coin (1q-coin) and two-qubit coin (2q-coin) QWs on a $2^n$-node ring for $2 \geq n \geq 4$, using the gate circuits in Appendix A. Gate errors, passive noise, state preparation, and readout are modelled as described in Section \ref{neutral-atoms}. For simplicity, the coin parameters are static with $\theta_t=\phi_t=\frac{\pi}{2}$ for all $t$. In this section, we consider `near-term' implementations and therefore use the native gate set $\textbf{G}(\max(r)= 3) = \{\text{H}, \text{X}, \text{R}_y(\frac{\pi}{2}), \text{CZ}_\text{eff}, \text{CCZ}_\text{eff}\}$.

\begin{figure}[h!]
    \centering
    \includegraphics[width=0.7\linewidth]{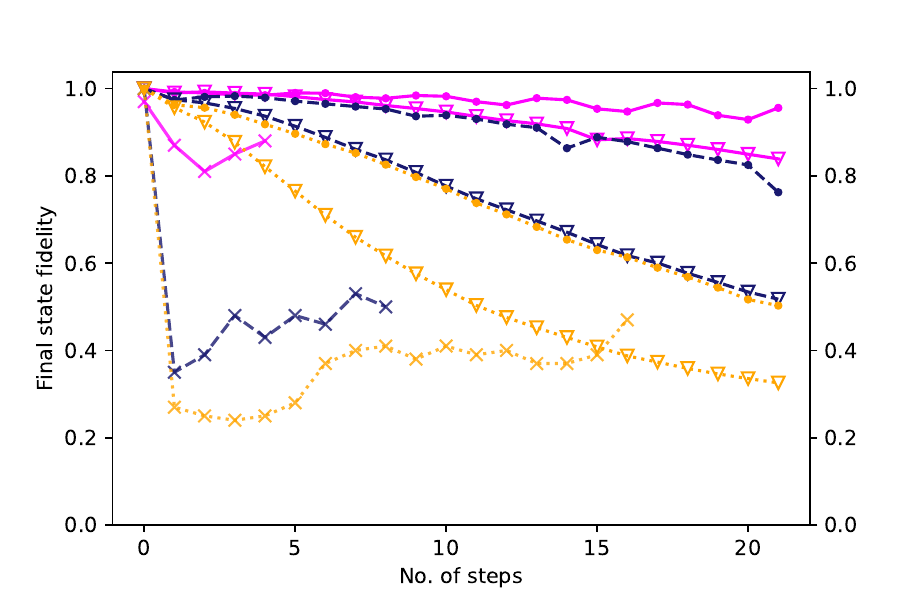}
    \caption{Final fidelity of the final position states between errorless and simulated error 1q-coin (circles) and 2q-coin (triangles) QWs on $2^n$-node rings, using native gate set $\textbf{G}(\max(r)= 3)$. Pink solid lines denote QWs on 4-node rings, dashed blue on 8-node rings, and dotted gold lines on 16-node rings. IBM results (using native gates of $\max(r)=2$) \cite{Wadhia_2024} for 1q-coin QWs are also shown for comparison in crosses.}
    \label{max3}
\end{figure}

Final state fidelity against number of steps are shown in Figure \ref{max3}, along with the IBM implementation in Wadhia et al. \cite{Wadhia_2024} for comparison. Note that only the results for the 1q-coin QWs are available for the IBM comparison.
It is clear that the overriding factor in state fidelity per step is the maximum rank of the gates in the QW gate circuit prior to decomposition, equivalent to the total number of position qubits and coin qubits: $n_q=n+1$ for 1q-coin QWs and $n_q=n+2$ for 2q-coin QWs. In general, the state fidelity decreases as $n_q$ increases. Further, the fidelity curves for QWs with equal $n_q$ values are roughly equivalent: 2q-coin QWs on a $2^n$-node ring results match those for 1q-coin QWs on a $2^{n+1}$-node ring.

The gate errors contribute overwhelmingly to state fidelity loss -- with only gate errors considered, the increase in fidelity is only 0.77\% - 1.76\% for the 1q-coin QW on a 4-node ring, which can almost entirely be attributed to atom loss at readout. This also results in the wobbles in fidelity at certain timesteps, which are not related to the number of steps but are artifacts of the randomness of the atom loss.

We find the number of steps for which each walk is within final state fidelity tolerances $\{0.9, 0.99\}$, shown in Table \ref{tol3}. Only the QWs on a 4-node ring have final state fidelity within 0.99 after any number of steps. The number of steps within tolerance decreases with both number of nodes $2^n$ and number of coin qubits $n_c$, with $n_q = n + n_c$ as the leading factor.
(Note that the state fidelity as a function of the number of steps is subpolynomial for the first few steps, which is much more noticeable for the 2q-coin QW on a 16-node ring.)

\begin{table}[h!]
\centering
 \begin{tabular}{c | c c} 
 QW & 0.9 & 0.99 \\ [0.5ex] 
 \hline
 1q-coin on 4-node & $\geq21$ & 2 \\ 
 2q-coin on 4-node & 14 & 2 \\
 1q-coin on 8-node & 13 & 0 \\
 2q-coin on 8-node & 5 & 0 \\
 1q-coin on 16-node & 4 & 0 \\
 2q-coin on 16-node & 2 & 0 \\ [1ex] 
 \end{tabular}
 \caption{Number of steps for which each QW is within final state fidelity tolerances $\{0.9, 0.99\}$.}
 \label{tol3}
\end{table}

Therefore of those considered, the only QWs feasible -- which here we mean which have state fidelity within tolerance 0.99 for at least two steps -- are 1q- or 2q-coin QWs on a 4-node ring. 1q-coin QWs on a 4-node ring have the added benefit of maintaining state fidelity above 0.9 for at least 21 steps.

\subsection{Further term}

We wish to identify the best potential hardware improvements for the application on 1q-coin and 2q-coin QWs.

\begin{figure}[h!]
    \centering
    \includegraphics[width=0.7\linewidth]{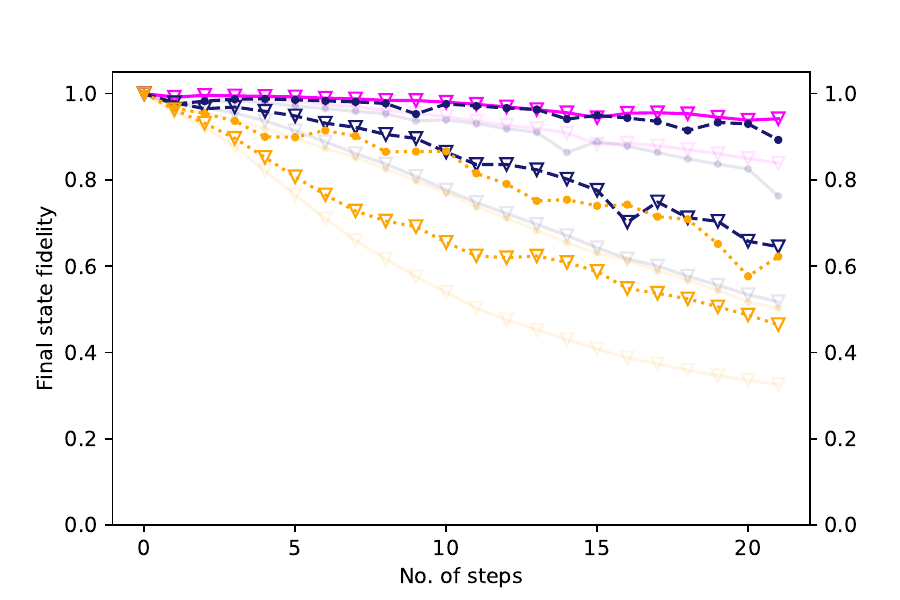}
    \caption{Final fidelity of the position states between errorless and simulated error QWs with native gate set $\textbf{G}(\max(r)= 4)$. Circles denote 1q-coin QWs, and triangles 2q-coin QWs. Pink solid lines are on 4-node ring, dashed blue on 8-node rings, and dotted gold lines on 16-node rings. In low opacity are the results of QWs with native gate set $\textbf{G}(\max(r)= 3)$ for comparison.}
    \label{fidsmax4}
\end{figure}

First we incorporate the use of the 4-qubit gate $\text{C}3\text{Z}$ with the effective matrix from equation \ref{c3zeff}, expanding our native gate set to $\textbf{G}(\max(r)=4) = \{\text{H}, \text{X}, \text{R}_y(\frac{\pi}{2}), \text{CZ}_\text{eff}, \text{CCZ}_\text{eff}, \text{C3Z}_\text{eff}\}$.
Figure \ref{fidsmax4} shows the final state fidelities for the considered QWs with this gate set, along with the previous section's results with gate set $\textbf{G}(r \leq 3)$ in reduced opacity for comparison. With native gate set $\textbf{G}(\max(r)=3)$ (from Section \ref{near-term}), after 21 steps of a 2q-coin QW on a 4-node ring, the final state fidelity is $f=0.84$. With native gate set $\textbf{G}(\max(r)=4)$, after 21 steps of the same size QW the final state fidelity is $f=0.94$, representing a $12\%$ increase.
Due to the greater asymmetry in $\text{C}3\text{Z}_\text{eff}$ as opposed to the other effective matrices (see Equations \ref{czeff}-\ref{c3zeff}), the fidelity curves are much less smooth. (This can be verified by constructing and applying symmetric effective matrices $\text{C}k\text{Z}_\text{eff} = \ketbra{0^{\otimes(k+1)}}{0^{\otimes(k+1)}} + \alpha e^{\beta i\pi}\sum_{\sigma \in \{0,1\}^{\otimes(k+1)}\backslash\{0\}^{\otimes(k+1)}}\ketbra{\sigma}{\sigma}$, which result in smooth final state fidelity curves.)

At large ($>10$) numbers of steps the improvement in state fidelity is significant: the number of steps within final state fidelity tolerance 0.9 for QWs with native gate set $\textbf{G}(r\leq4)$ increase to $\{21,20,8\}$ for QW $\in \{$2q-coin on 4-node, 1q-coin on 8-node, 2q-coin on 8-node$\}$ (from $\{14,13,4\}$ with $\textbf{G}(r\leq3)$). However, for few steps there is only a little improvement, with the number of steps within tolerance 0.99 increasing from 2 to 5 for a 2q-coin QW on a 4-node ring. Otherwise there are no step increases for tolerances $\{0.9, 0.99, 0.999\}$.
%0.9: [], 21, 20, 8, 3?, 2

%CCZdiag = [fid * np.exp(fid * 1j * np.pi)] * 2**3
Another route is to improve the gate fidelities of CZ and CCZ. In order to model further term gate fidelities we construct effective matrices
\begin{align}
    \text{C}\text{Z}(a) = \text{diag}\left[1, \;\alpha_1(a) e^{i \pi\phi_1(a)}, \; \alpha_1(a) e^{i \pi\phi_1(a)}, \; \frac{\alpha_2(a=0)}{\alpha_1(a=0)}\alpha_1(a) e^{i \pi\frac{\phi_2(a=0)}{\phi_1(a=0)}\phi_1(a)}\right]
\end{align}
where $\alpha_1(a) = 0.9990 + 0.0001a \leq 1$, $\phi_1(a) = 0.9906 + 0.0010a \leq 1$, and $\text{CZ}(a=0)=\text{CZ}_\text{eff}$. Therefore the ratio between gate coefficients is maintained. We similarly construct $\text{CC}\text{Z}(a)$, and only consider gate pairs $[\text{CZ}(a), \text{CCZ}(a)]$.%, with fidelities
%\begin{align}
    %\mathcal{F}(\text{C}k\text{Z})= \left(\frac{1}{2^{k+1}}\right)^2 \left|1 -  \sum_j^{k+1} A_j e^{i\pi\phi_j}\right|^2 .
%\end{align}

\begin{figure}[h!]
    \begin{subfigure}{0.49\textwidth}
        \includegraphics[width=\linewidth]{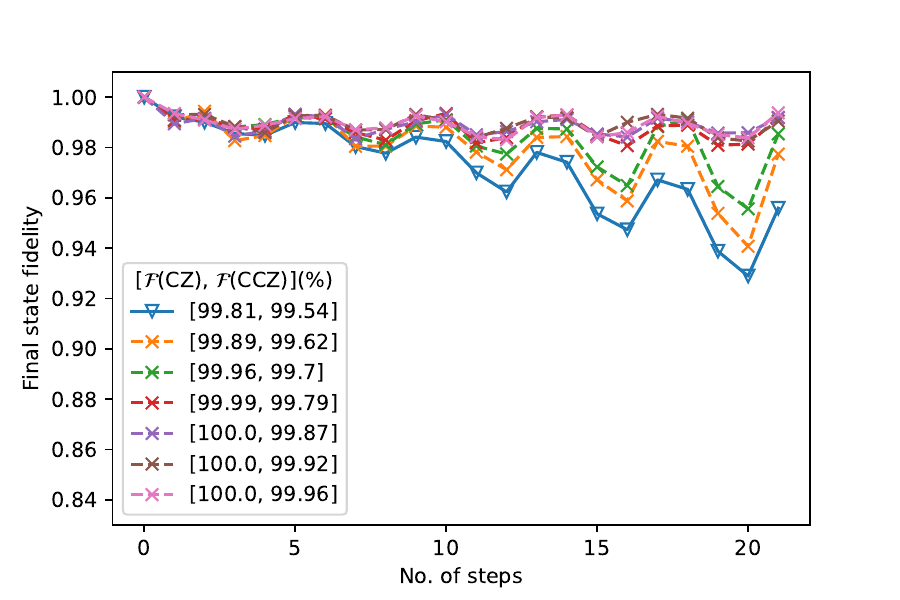}
        \caption{1q-coin QW on a 4-node ring ($n_q=3$ qubits).}
    \end{subfigure}
    \begin{subfigure}{0.49\textwidth}
        \includegraphics[width=\linewidth]{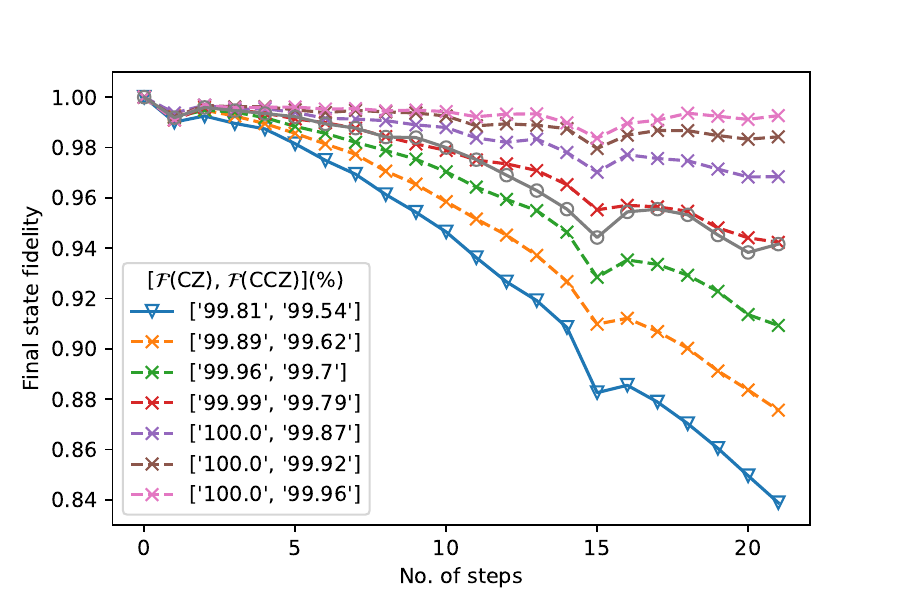}
        \caption{2q-coin QW on a 4-node ring ($n_q=4$ qubits).}
    \end{subfigure}
    \caption{Final state fidelity at each step QWs on a 4-node ring using native gate set $\textbf{G}(\max(r)= 3)(a)$ with various gate fidelities $\mathcal{F}(\text{CCZ})$ and $\mathcal{F}(\text{CZ})$. Also shown, for the 2q-coin QW, are the results using native gate set $\textbf{G}(\max(r)= 4)$ in circles for comparison.}
    \label{fids}
\end{figure}

Figure \ref{fids} shows the final state fidelities for 1q-coin and 2q-coin QWs on a 4-node ring with native gate sets $\textbf{G}(\max(r)= 3)(a) = \{\text{H}, \text{X}, \text{R}_y(\frac{\pi}{2}), \text{CZ}(a), \text{CCZ}(a)\}$ where $a \in \{0, \frac{13}{3}, \frac{26}{3}, 13, \frac{52}{3}, \frac{65}{3}, 26 \}$. Note that the troughs in the state fidelity curves are due to phase values $\phi_j(a)\neq1$, however they become less significant as the coefficients $\alpha_j(a)$ approach unity. The sudden jump in smoothness found with gate fidelities $[\mathcal{F} (\text{CZ}),\mathcal{F} (\text{CCZ})]=[99.99, 99.79]\%$ corresponds to $\alpha_1(a) = 1$.

There is no change in the tolerance results for the 1q-coin QW on a 4-node ring for tolerances $\{0.9, 0.99, 0.999\}$. The improvements are more apparent in the other walks, for which the number of steps within tolerance 0.9 improve steadily with gate fidelity, however none improve their number of steps within tolerance 0.99.

On Figure \ref{fids}(b) we have also plotted the 
$\textbf{G}(\max(r)= 4)$ gate set result for comparison: we can see that it roughly aligns with the gate set $\textbf{G}(\max(r)= 3)(a=13)$ with $\mathcal{F} (\text{CZ})=99.99\%$ and $\mathcal{F} (\text{CCZ})=99.79\%$. These CZ and CCZ gate fidelities are also required for the number of steps within tolerance results to match. Similarly, for a 1q-coin QW on an 8-node ring, the state fidelity curve roughly matches that with $\mathcal{F} (\text{CZ})=99.98\%$ and $\mathcal{F} (\text{CCZ})= 99.75\%$. The fidelity pairs required continue to decrease to $\mathcal{F} (\text{CZ})=99.96\%$ and $\mathcal{F} (\text{CCZ})= 99.70\%$ for 2q-coin QW on 8-node and 1q-coin QW on 16-node. This suggests that while four-qubit gates are hugely beneficial to quantum walks on small rings, they may not have as marked an effect on quantum walks on very large rings.

%% file: content/conc.tex
\section{Conclusions}\label{sec:conc}
We show that with projected gate fidelities of $\mathcal{F}(\text{CZ}) = 0.9981$ and $\mathcal{F}(\text{CCZ}) = 0.9954$, a quantum walk with a rest-state performed on a 4-node ring (requiring four qubits) results in final state fidelity $f>0.99$ after two time steps and $f>0.90$ after fourteen timesteps. We find that these can be increased to five timesteps with $f>0.99$ and at least twenty timesteps with $f>0.90$ if either the CZ and CCZ gate fidelities are increased to $\mathcal{F} (\text{CZ})=99.99\%$ and $\mathcal{F} (\text{CCZ})=99.79\%$ or, perhaps more simply, we introduce a native four-qubit gate with $\mathcal{F}(\text{C3Z}) = 0.9850$. With these and similar results for quantum walks on 8- and 16-node rings, we therefore conclude that native four-qubit gates and mid-circuit rearrangement is highly recommended for the near-term implementation of interesting toy quantum walks on neutral atom hardware.

%The increase in state fidelity with the use of a native four-qubit gate compared to improved two- and three-qubit gate fidelity decrease slightly with larger rings. This suggests that while native four-qubit gates and mid-circuit rearrangement is imperative to the near-term implementation of interesting quantum walks on neutral atom hardware, this may not be an ongoing requirement for quantum walks on large rings in future fault tolerant hardware.

\begin{figure}[h!]
    \centering
    \includegraphics[width=.7\linewidth]{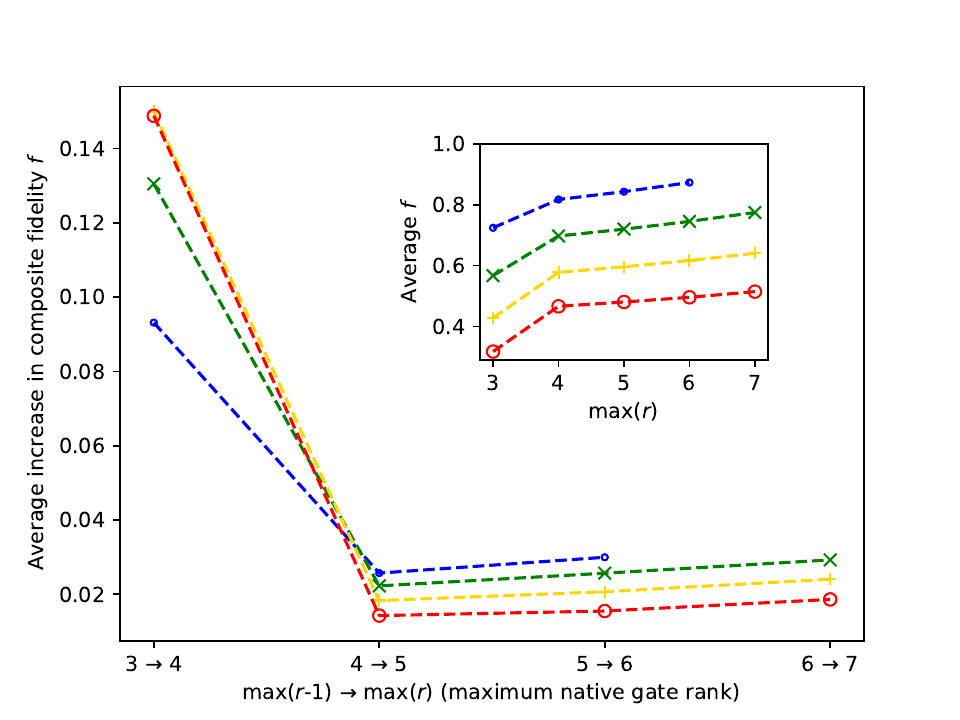}
    \caption{Mean increase in composite fidelity $f(\max(r))=\prod_{k=2}^{\max(r)-1} \mathcal{F}(\text{C}k\text{Z})^{\text{no. of }\text{C}k\text{Z}}$ from using native gate set $\textbf{G}(\text{max}(r))$ compared to using gate set $\textbf{G}(\text{max}(r)-1)$, with the number of multiqubit gates required for one step of a 2q-coin QW on a $2^n$-node ring, where $n \in \{ 6, 8, 10, 12\}$ in blue dots, green crosses, gold pluses, and red circles respectively. Inset shows corresponding average composite fidelity.}
    %=\{\text{C}\text{Z}\}
    \label{maxg}
\end{figure}

As we look to the further term and the goal of implementing useful QWs (requiring more than five qubits), this raises the concern that QWs on $2^n$-node ring require native gate ranks (number of qubits acted on) of at least $n+1$. We perform a back of the envelope calculation to find composite fidelities $f(\max(r))=\prod_{k=2}^{\max(r)-1} \mathcal{F}(\text{C}k\text{Z})^{\text{no. of }\text{C}k\text{Z}}$ for a two-qubit coin quantum walk on a $2^{n}$-node ring for various $n$ and various native gate sets $\textbf{G}(\max(r))=\{\text{C}k\text{Z} | 2 \leq k < \max(r)\}$, shown in Figure \ref{maxg}, where $\max(r)$ is the maximum rank of the gates in the gate set. 
Gate fidelities considered are such that $0.9954 \leq \mathcal{F}(\text{CCZ}) \leq 0.9999$ and $0.9960 \leq \mathcal{F}(\text{C}k\text{Z})/\mathcal{F}(\text{C}(k-1)\text{Z}) \leq 1$, chosen so that the average increase in composite fidelity is always positive. There is a clear advantage in increasing the gate set from $\textbf{G}(\max(r)= 3)$ to $\textbf{G}(\max(r)= 4)$ -- this advantage however is not as pronounced for further increases in gate sets.
Therefore, we propose that while the introduction of C3Z gates would yield relatively large final state fidelity increases, it is not necessary to continue increasing native gate ranks ad infinitum, and indeed $\text{max}(r)=4$ seems to be the sweet spot. When realistic modelling for C4Z gates and greater ranks are obtained, this conjecture should be further investigated.
%\text{CC}\text{Z}, ..., \text{C}(\max(r)-1)\text{Z}

%... it raises concern for the scalability of 1D quantum walks; $2^n$-node requiring $\approx n$-qubit gates is not feasible for all $n$. We therefore briefly investigate far-term, large $2^n$-node QWs and their circuit fidelities with maximum decomposition gate sizes of 4, 5, and 6.

%NO EFF MATRICES FOR C4Z SO NO SIMULATION, THESE ARE ROUGH ESTIMATES USING COMPOSITE FIDELITY $f=\sum_{k=2}^{d-1} \mathcal{F}(\text{C}k\text{Z})^{\text{no. of }\text{C}k\text{Z}}$.

%Figure \ref{345} shows an estimate of the combined fidelities after one step using calculated [gerard] CCZ and C3Z gate fidelities and a range of far-term C4Z fidelities for some mid-sized ($5 \leq s \leq 7$) QWs: this shows a gate fidelity of $\mathcal{F}(\text{C}4\text{Z}) \geq 0.9750$ is required for for improvement over $\mathcal{F}(\text{C}3\text{Z}) = 0.9850$. Figure \ref{345far} shows that for max decomp size 3, a gate fidelity of $\mathcal{F}(\text{CC}\text{Z}) \approx 0.99999$ is required to beat fidelities of $\mathcal{F}(\text{C}(k={3,4,5})\text{Z}) \approx 0.99990$ for any max decomp size $\in {4,5,6}$. However, within $4 \leq \text{maxd} \leq 6$, max decomp size plays a relatively small role on the composite fidelity. 

Further work can be split into theory and experiment. For the theoretical implementation, position encoding with Gray codes should be considered, to potentially decrease the gate requirements. Only qubits were considered; the use of qutrit coins could also decrease gate requirements, although careful consideration of qutrit gate operation would be required. No error correction was considered; with the advent of fault-tolerant protocols for multiqubit gates \cite{old2025faulttolerantstabilizermeasurementssurface}, this should be added to the error model.
With the specific view of the application to fluid simulation, we wish to extend the linear case considered in this work to the non-linear case by entangling the coin qubit(s) with an additional ancilla. Further term, we look to the extension of quantum walks on 1D rings to those on 2D lattices.

For experimentalists, we would like to see a quantum walk performed on a 4-node ring in order to compare our theoretical results.

%% file: content/appendix.tex
\newpage
\appendix
\section*{APPENDIX}
 
Here we present the explicit gate circuit decompositions used in the detailed realistic simulations. We also denote where qubit movement is required (in a 2D array with maximum static connectivity of four qubits per qubit) with the single qubit gate `M' (see Section \ref{neutral-atoms} for details). In Appendix \ref{app:circ3} we present the circuits with native gate set
$\textbf{G}(\max(r)=3) = \{\text{H}, \text{X}, \text{R}_y(\frac{\pi}{2}), \text{CZ}_\text{eff}, \text{CCZ}_\text{eff}\}$ and in Appendix \ref{app:circ4} those with $\textbf{G}(\max(r)=4) = \{\text{H}, \text{X}, \text{R}_y(\frac{\pi}{2}), \text{CZ}_\text{eff}, \text{CCZ}_\text{eff}, \text{C3Z}_\text{eff}\}$. For brevity we do not write out the transformation C$k$X $\rightarrow$ C$k$Z (Equation \ref{CZtoCX}) explicitly.

%%%%%%%%%%%%%%%%%%%%%%%%%%%%%%%%%%%%%%%%%%%%%%%%%%%%%%%%%%%%%%%%
\section{QW circuits with native gates of maximum rank 3} \label{app:circ3}
%%%%%%%%%%%%%%%%%%%%%%%%%%%%%%%%%%
\subsection{1q-coin QWs}

% NL 4-node
\begin{figure}[h!]
    \centering
    \begin{tikzpicture}[thick]
        \tikzset{
        operator/.style = {draw,fill=white,minimum size=0.1em}, 
        operator2/.style = {draw,fill=white,minimum height=2cm, minimum width=1cm}, 
        phase/.style = {draw,fill,shape=circle,minimum size=5pt,inner sep=0pt}, 
        surround/.style = {fill=blue!10,thick,draw=black,rounded corners=2mm}, 
        cross/.style={path picture={\draw[thick,black](path picture bounding box.north) -- (path picture bounding box.south) (path picture bounding box.west) -- (path picture bounding box.east); }}, crossx/.style={path picture={ \draw[thick,black,inner sep=0pt] (path picture bounding box.south east) -- (path picture bounding box.north west) (path picture bounding box.south west) -- (path picture bounding box.north east); }}, 
        circlewc/.style={draw,circle,cross,minimum width=0.3 cm}, 
        meter/.style= {draw, fill=white, inner sep=5, rectangle, font=\vphantom{A}, minimum width=20, line width=.8, path picture={\draw[black] ([shift={(.1,.2)}]path picture bounding box.south west) to[bend left=30] ([shift={(-.1,.2)}]path picture bounding box.south east);\draw[black,-latex] ([shift={(0,.1)}]path picture bounding box.south) -- ([shift={(.2,-.1)}]path picture bounding box.north);}}, }
        \matrix[row sep=0.2cm, column sep=0.35cm] (circuit) {
        \node (q1) {$x_1$: $\ket{0}$}; % LINE p1
        &&& \node[circlewc] (C1) {}; &&&&&\node[circlewc] (C2) {}; &&& &\coordinate (end1); &[-0.2cm]\node (d1) {$\cdots$}; &[-0.3cm]\node[meter] (meter) {};
        \\ \node (q2) {$x_2$: $\ket{0}$}; % LINE p2
        &&&\node[phase] (P0) {}; && \node[circlewc] (C21) {}; &&\node[operator] (H0) {X}; &\node[phase] (P0) {}; &\node[operator] (H0) {X}; &\node[circlewc] (C22) {}; && \coordinate (end2); &[-0.2cm]\node (d1) {$\cdots$}; &[-0.3cm]\node[meter] (meter) {};
        \\ \node (q3) {$c_1$: $\ket{0}$}; % LINE c1
        &\node[operator] (H11) {$\text{R}_y(\theta_t)$}; &&\node[phase] (P2) {}; &&\node[phase] (P1) {}; &&\node[operator] (X1) {X}; &\node[phase] (P11) {}; &&\node[phase] (P3) {}; &\node[operator] (H0) {X}; & \coordinate (end3); &[-0.2cm]\node (d1) {$\cdots$}; \\
        && \node (l1dn) {}; &&&& \node (l2dn) {};  % LINE 0dn
        \\ 
        };
        \begin{pgfonlayer}{background} 
        \draw[thick] (q1) -- (end1) (q2) -- (end2) (q3) -- (end3) (P1) -- (C21) (P2) -- (C1) (P3) -- (C22) (P11) -- (C2);
        \end{pgfonlayer}
    \end{tikzpicture}
    \caption{1q-coin QW on a 4-node ring with native gate set $\textbf{G}(\text{max}(r) = 3)$.}
\end{figure}

% max3 NL 8-node
\begin{figure}[h!]
    \centering
    \begin{tikzpicture}[thick]
        \tikzset{
        operator/.style = {draw,fill=white,minimum size=0.1em}, 
        operator2/.style = {draw,fill=white,minimum height=2cm, minimum width=1cm}, 
        phase/.style = {draw,fill,shape=circle,minimum size=5pt,inner sep=0pt}, 
        surround/.style = {fill=blue!10,thick,draw=black,rounded corners=2mm}, 
        cross/.style={path picture={\draw[thick,black](path picture bounding box.north) -- (path picture bounding box.south) (path picture bounding box.west) -- (path picture bounding box.east); }}, crossx/.style={path picture={ \draw[thick,black,inner sep=0pt] (path picture bounding box.south east) -- (path picture bounding box.north west) (path picture bounding box.south west) -- (path picture bounding box.north east); }}, 
        circlewc/.style={draw,circle,cross,minimum width=0.3 cm}, 
        meter/.style= {draw, fill=white, inner sep=5, rectangle, font=\vphantom{A}, minimum width=20, line width=.8, path picture={\draw[black] ([shift={(.1,.2)}]path picture bounding box.south west) to[bend left=30] ([shift={(-.1,.2)}]path picture bounding box.south east);\draw[black,-latex] ([shift={(0,.1)}]path picture bounding box.south) -- ([shift={(.2,-.1)}]path picture bounding box.north);}}, }
        \matrix[row sep=0.2cm, column sep=0.35cm] (circuit) {
        \node (q1) {$x_1$: $\ket{0}$}; % LINE x1
        & & &\node[circlewc] (C11) {}; & &\node[circlewc] (C12) {}; &&&&&& &\node[circlewc] (C13) {}; & &\node[circlewc] (C14) {}; &&&&& &\coordinate (end1); &[-0.2cm]\node (d1) {$\cdots$}; &[-0.3cm]\node[meter] (meter) {};
        \\ \node (q2) {$x_2$: $\ket{0}$}; % LINE x2
        & & &\node[phase] (P21) {}; & &\node[phase] (P22) {}; &&\node[circlewc] (C21) {}; &&&\node[operator] (X11) {X}; & &\node[phase] (P23) {}; & &\node[phase] (P24) {}; &\node[operator] (X12) {X}; &\node[circlewc] (C22) {}; &&& &\coordinate (end2); &[-0.2cm]\node (d2) {$\cdots$}; &[-0.3cm]\node[meter] (meter) {};
        \\ \node (q3) {$x_3$: $\ket{0}$}; % LINE x3
        & &\node[phase] (P31) {}; & &\node[phase] (P32) {}; &&&\node[phase] (P33) {}; &&\node[circlewc] (C31) {}; &\node[operator] (X21) {X}; &\node[phase] (P34) {}; & &\node[phase] (P35) {};
        &&&\node[phase] (P36) {};
        &\node[operator] (X22) {X}; &\node[circlewc] (C32) {}; & &\coordinate (end3); &[-0.2cm]\node (d3) {$\cdots$}; &[-0.3cm]\node[meter] (meter) {};
        \\ \node (q4) {$c_1$: $\ket{0}$}; % LINE c1
        &\node[operator] (H11) {$\text{R}_y(\theta_t)$}; &\node[phase] (P11) {}; &&\node[phase] (P12) {};
        &&\node[operator] (M1) {M};
        &\node[phase] (P13) {};
        &\node[operator] (M1) {M};
        &\node[phase] (P14) {};
        &\node[operator] (X31) {X}; &\node[phase] (P15) {};
        &&\node[phase] (P16) {};
        &&\node[operator] (M1) {M};
        &\node[phase] (P17) {};
        &\node[operator] (M1) {M};
        &\node[phase] (P18) {};
        &\node[operator] (X32) {X};
        &\coordinate (end4); &[-0.2cm]\node (d1) {$\cdots$};
        \\
        \node (q5) {$a_1$: $\ket{0}$}; % LINE a1
        & &\node[circlewc] (C51) {}; &\node[phase] (P51) {}; &\node[circlewc] (C52) {}; &\node[phase] (P52) {};
        &&&&& &\node[circlewc] (C53) {}; &\node[phase] (P53) {}; &\node[circlewc] (C54) {}; &\node[phase] (P54) {}; &&&&& &\coordinate (end5); &[-0.2cm]\node (d1) {$\cdots$};
        \\};
        \begin{pgfonlayer}{background} 
        \draw[thick] (q1) -- (end1) (q2) -- (end2) (q3) -- (end3) (q4) -- (end4) (q5) -- (end5)
        (P31) -- (C51) (P51) -- (C11) (P32) -- (C52) (P52) -- (C12)
        (C21) -- (P13) (C31) -- (P14)
        (P34) -- (C53) (P53) -- (C13) (P35) -- (C54) (P54) -- (C14)
        (P17) -- (C22) (P18) -- (C32)
        ; \end{pgfonlayer}
    \end{tikzpicture}
    \caption{1q-coin QW on a 8-node ring with native gate set $\textbf{G}(\text{max}(r) = 3)$.}
\end{figure}

% max3 NL 16-node
\begin{figure}[h!]
    \centering
    \begin{tikzpicture}[thick]
        \tikzset{
        operator/.style = {draw,fill=white,minimum size=0.05em}, 
        operator2/.style = {draw,fill=white,minimum height=2cm, minimum width=1cm}, 
        phase/.style = {draw,fill,shape=circle,minimum size=5pt,inner sep=0pt}, 
        surround/.style = {fill=blue!10,thick,draw=black,rounded corners=2mm}, 
        cross/.style={path picture={\draw[thick,black](path picture bounding box.north) -- (path picture bounding box.south) (path picture bounding box.west) -- (path picture bounding box.east); }}, crossx/.style={path picture={ \draw[thick,black,inner sep=0pt] (path picture bounding box.south east) -- (path picture bounding box.north west) (path picture bounding box.south west) -- (path picture bounding box.north east); }}, 
        circlewc/.style={draw,circle,cross,minimum width=0.3 cm}, 
        meter/.style= {draw, fill=white, inner sep=5, rectangle, font=\vphantom{A}, minimum width=20, line width=.8, path picture={\draw[black] ([shift={(.1,.2)}]path picture bounding box.south west) to[bend left=30] ([shift={(-.1,.2)}]path picture bounding box.south east);\draw[black,-latex] ([shift={(0,.1)}]path picture bounding box.south) -- ([shift={(.2,-.1)}]path picture bounding box.north);}}, }
        \matrix[row sep=0.2cm, column sep=0.3cm] (circuit) {
        &&&&& &&& &\node (up1) {}; &&&&&&& &\node (up2) {}; && &\node (up3) {}; & &\node (up4) {}; % top line
        \\\node (q1) {$x_1$:}; % LINE x1
        &\node[circlewc] (C11) {}; &&& &\node[circlewc] (C12) {}; &&&&&&&&&&&&&&&&&& &\coordinate (end1); &[0.05cm]\node (d1) {$\cdots$}; %&[0.05cm]\node[meter] (meter) {};
        \\ \node (q2) {$x_2$:}; % LINE x2
        &\node[phase] (P21) {}; &&& &\node[phase] (P22) {}; &&&&&& &\node[circlewc] (C21) {}; &&&
        &\node[circlewc] (C22) {}; &&&&&&&
        &\coordinate (end2); &[-0.2cm]\node (d2) {$\cdots$}; %&[-0.3cm]\node[meter] (meter) {};
        \\ \node (q3) {$x_3$:}; % LINE x3
        & &\node[phase] (P31) {}; & &\node[phase] (P32) {}; & &\node[phase] (P33) {}; & &\node[phase] (P34) {}; &&& &\node[phase] (P35) {}; &&&
        &\node[phase] (P36) {}; && &\node[circlewc] (C31) {};
        &&&&&\coordinate (end3); &[-0.2cm]\node (d2) {$\cdots$}; %&[-0.3cm]\node[meter] (meter) {};
        \\ \node (q4) {$x_4$:}; % LINE x4
        && &\node[phase] (Pc11) {}; &&&
        &\node[phase] (Pc12) {}; &&
        &\node[phase] (Pc13) {}; &&&
        &\node[phase] (Pc14) {}; &&&&
        &\node[phase] (Pc15) {}; &
        &\node[circlewc] (Cend) {};
        &&&\coordinate (end4); &[-0.2cm]\node (d1) {$\cdots$};
        \\ \node (q5) {$c_1$:}; % LINE c1
        && &\node[phase] (Pc21) {}; &&&
        &\node[phase] (Pc22) {}; &&
        &\node[phase] (Pc23) {}; &&&
        &\node[phase] (Pc24) {}; &&&
        &\node[operator] (M1) {M};
        &\node[phase] (Pc25) {}; 
        &&\node[phase] (Pend) {};
        &&\node[operator] (M2) {M};
        &\coordinate (end5); &[-0.2cm]\node (d1) {$\cdots$};
        \\
        \node (q6) {$a_1$:}; % LINE a1
        & &\node[phase] (Pa11) {}; &\node[circlewc] (Ca11) {};  &\node[phase] (Pa12) {}; &
        &\node[phase] (Pa13) {}; &\node[circlewc] (Ca12) {}; &\node[phase] (Pa14) {}; &
        &\node[circlewc] (Ca13) {}; 
        &\node[operator] (M1) {M};
        &\node[phase] (Pa15) {};
        &\node[operator] (M2) {M};
        &\node[circlewc] (Ca14) {};
        &\node[operator] (M3) {M};
        &\node[phase] (Pa16) {};
        &&&&&&&\node[operator] (M4) {M};
        &\coordinate (end6); &[-0.2cm]\node (d1) {$\cdots$};
        \\ \node (q7) {$a_2$:}; % LINE a2
        &\node[phase] (Pa21) {}; &\node[circlewc] (Ca21) {}; &
        &\node[circlewc] (Ca22) {};
        &\node[phase] (Pa22) {}; 
        &\node[circlewc] (Ca23) {}; &
        &\node[circlewc] (Ca24) {}; &&&&&&&&&&&&& 
        &&
        &\coordinate (end7); &[0.2cm]\node (d1) {$\cdots$};
        \\
        &&&&& &&& &\node (down1) {}; &&&&&&& &\node (down2) {}; && &\node (down3) {}; & &\node (down4) {}; % bottom line
        \\};
        \begin{pgfonlayer}{background} 
        \draw[thick] (q1) -- (end1) (q2) -- (end2) (q3) -- (end3) (q4) -- (end4) (q5) -- (end5) (q6) -- (end6) (q7) -- (end7)
        (Pc11) -- (Ca11)
        (P31) -- (Ca21) (Pa21) -- (C11)
        (P32) -- (Ca22) (Pa22) -- (C12)
        (Pc12) -- (Ca12) (Cend) -- (Pend)
        (P33) -- (Ca23)
        (P34) -- (Ca24)
        (Pc13) -- (Ca13)
        (Pa15) -- (C21)
        (Pc14) -- (Ca14)
        (Pa16) -- (C22)
        (C31) -- (Pc25);
        \draw[thick, dotted] (up1) -- (down1) (up2) -- (down2) (up3) -- (down3) (up4) -- (down4)
        ; \end{pgfonlayer}
    \end{tikzpicture}
    \caption{1q-coin QW on a 16-node ring with native gate set $\textbf{G}(\text{max}(r) = 3)$;  \textbf{increment only}.}
\end{figure}

%%%%%%%%%%%%%%%%%%%%%%%%%%%%%%%%%%%%
\newpage
\subsection{2q-coin QWs}

% max3 Lazy 4-node
\begin{figure}[h!]
    \centering
    \begin{tikzpicture}[thick]
        \tikzset{
        operator/.style = {draw,fill=white,minimum size=0.1em}, 
        operator2/.style = {draw,fill=white,minimum height=2cm, minimum width=1cm}, 
        phase/.style = {draw,fill,shape=circle,minimum size=5pt,inner sep=0pt}, 
        surround/.style = {fill=blue!10,thick,draw=black,rounded corners=2mm}, 
        cross/.style={path picture={\draw[thick,black](path picture bounding box.north) -- (path picture bounding box.south) (path picture bounding box.west) -- (path picture bounding box.east); }}, crossx/.style={path picture={ \draw[thick,black,inner sep=0pt] (path picture bounding box.south east) -- (path picture bounding box.north west) (path picture bounding box.south west) -- (path picture bounding box.north east); }}, 
        circlewc/.style={draw,circle,cross,minimum width=0.3 cm}, 
        meter/.style= {draw, fill=white, inner sep=5, rectangle, font=\vphantom{A}, minimum width=20, line width=.8, path picture={\draw[black] ([shift={(.1,.2)}]path picture bounding box.south west) to[bend left=30] ([shift={(-.1,.2)}]path picture bounding box.south east);\draw[black,-latex] ([shift={(0,.1)}]path picture bounding box.south) -- ([shift={(.2,-.1)}]path picture bounding box.north);}}, }
        \matrix[row sep=0.2cm, column sep=0.35cm] (circuit) {
        \node (q1) {$x_1$: $\ket{0}$}; % LINE p1
        & & &\node[circlewc] (C11) {}; & &\node[circlewc] (C12) {}; &&&&& &\node[circlewc] (C13) {}; & &\node[circlewc] (C14) {}; &&& &\coordinate (end1); &[-0.2cm]\node (d1) {$\cdots$}; &[-0.3cm]\node[meter] (meter) {};
        \\ \node (q2) {$x_2$: $\ket{0}$}; % LINE p2
        & & &\node[phase] (P21) {}; & &\node[phase] (P22) {}; &&\node[circlewc] (C21) {}; &&\node[operator] (X11) {X}; & &\node[phase] (P23) {}; & &\node[phase] (P24) {}; &\node[operator] (X12) {X}; &\node[circlewc] (C22) {}; & &\coordinate (end2); &[-0.2cm]\node (d2) {$\cdots$}; &[-0.3cm]\node[meter] (meter) {};
        \\ \node (q3) {$c_1$: $\ket{0}$}; % LINE c1
        &\node[operator] (H11) {$\text{R}_y(\theta_t)$}; &\node[phase] (P31) {}; & &\node[phase] (P32) {}; &&&\node[phase] (P33) {}; &&\node[operator] (X21) {X}; &\node[phase] (P34) {}; & &\node[phase] (P35) {};
        &&&\node[phase] (P36) {};
        &\node[operator] (X22) {X}; &\coordinate (end3); &[-0.2cm]\node (d3) {$\cdots$}; &[-0.3cm]\node[meter] (meter) {};
        \\ \node (q4) {$c_2$: $\ket{0}$}; % LINE c2
        &\node[operator] (H11) {$\text{R}_y(\phi_t)$}; &\node[phase] (P11) {}; &&\node[phase] (P12) {};
        &&\node[operator] (M1) {M};
        &\node[phase] (P13) {};
        &\node[operator] (M2) {M}; &\node[operator] (X31) {X}; &\node[phase] (P15) {};
        &&\node[phase] (P16) {};
        &&\node[operator] (M3) {M};
        &\node[phase] (P17) {};
        &\node[operator] (M4) {M};
        &\coordinate (end4); &[-0.2cm]\node (d1) {$\cdots$};
        \\
        \node (q5) {$a_1$: $\ket{0}$}; % LINE a1
        & &\node[circlewc] (C51) {}; &\node[phase] (P51) {}; &\node[circlewc] (C52) {}; &\node[phase] (P52) {};
        &&&& &\node[circlewc] (C53) {}; &\node[phase] (P53) {}; &\node[circlewc] (C54) {}; &\node[phase] (P54) {}; &&& &\coordinate (end5); &[-0.2cm]\node (d1) {$\cdots$};
        \\};
        \begin{pgfonlayer}{background} 
        \draw[thick] (q1) -- (end1) (q2) -- (end2) (q3) -- (end3) (q4) -- (end4) (q5) -- (end5)
        (P31) -- (C51) (P51) -- (C11) (P32) -- (C52) (P52) -- (C12)
        (C21) -- (P13)
        (P34) -- (C53) (P53) -- (C13) (P35) -- (C54) (P54) -- (C14)
        (P17) -- (C22)
        ; \end{pgfonlayer}
    \end{tikzpicture}
    \caption{2q-coin QW on a 4-node ring with native gate set $\textbf{G}(\text{max}(r) = 3)$.}
\end{figure}

% max3 Lazy 8-node
\begin{figure}[h!]
    \centering
    \begin{tikzpicture}[thick]
        \tikzset{
        operator/.style = {draw,fill=white,minimum size=0.05em}, 
        operator2/.style = {draw,fill=white,minimum height=2cm, minimum width=1cm}, 
        phase/.style = {draw,fill,shape=circle,minimum size=5pt,inner sep=0pt}, 
        surround/.style = {fill=blue!10,thick,draw=black,rounded corners=2mm}, 
        cross/.style={path picture={\draw[thick,black](path picture bounding box.north) -- (path picture bounding box.south) (path picture bounding box.west) -- (path picture bounding box.east); }}, crossx/.style={path picture={ \draw[thick,black,inner sep=0pt] (path picture bounding box.south east) -- (path picture bounding box.north west) (path picture bounding box.south west) -- (path picture bounding box.north east); }}, 
        circlewc/.style={draw,circle,cross,minimum width=0.3 cm}, 
        meter/.style= {draw, fill=white, inner sep=5, rectangle, font=\vphantom{A}, minimum width=20, line width=.8, path picture={\draw[black] ([shift={(.1,.2)}]path picture bounding box.south west) to[bend left=30] ([shift={(-.1,.2)}]path picture bounding box.south east);\draw[black,-latex] ([shift={(0,.1)}]path picture bounding box.south) -- ([shift={(.2,-.1)}]path picture bounding box.north);}}, }
        \matrix[row sep=0.2cm, column sep=0.35cm] (circuit) {
        &&&&& &&& &\node (up2) {}; &&&&&&& &\node (up3) {}; && &\node (up4) {}; % top line
        \\\node (q1) {$x_1$:}; % LINE x1
        &\node[circlewc] (C11) {}; &&& &\node[circlewc] (C12) {}; &&&&&&&&&&&&&&&& &\coordinate (end1); &[0.05cm]\node (d1) {$\cdots$}; %&[0.05cm]\node[meter] (meter) {};
        \\ \node (q2) {$x_2$:}; % LINE x2
        &\node[phase] (P21) {}; &&& &\node[phase] (P22) {}; &&&&&& &\node[circlewc] (C21) {}; &&&
        &\node[circlewc] (C22) {}; &&&&&
        &\coordinate (end2); &[-0.2cm]\node (d2) {$\cdots$}; %&[-0.3cm]\node[meter] (meter) {};
        \\ \node (q3) {$x_3$:}; % LINE x3
        & &\node[phase] (P31) {}; & &\node[phase] (P32) {}; & &\node[phase] (P33) {}; & &\node[phase] (P34) {}; &&& &\node[phase] (P35) {}; &&&
        &\node[phase] (P36) {}; && &\node[circlewc] (C31) {};
        &&&\coordinate (end3); &[-0.2cm]\node (d2) {$\cdots$}; %&[-0.3cm]\node[meter] (meter) {};
        \\ \node (q4) {$c_1$:}; % LINE c1
        && &\node[phase] (Pc11) {}; &&&
        &\node[phase] (Pc12) {}; &&
        &\node[phase] (Pc13) {}; &&&
        &\node[phase] (Pc14) {}; &&&&
        &\node[phase] (Pc15) {};
        &&&\coordinate (end4); &[-0.2cm]\node (d1) {$\cdots$};
        \\ \node (q5) {$c_2$:}; % LINE c2
        && &\node[phase] (Pc21) {}; &&&
        &\node[phase] (Pc22) {}; &&
        &\node[phase] (Pc23) {}; &&&
        &\node[phase] (Pc24) {}; &&&
        &\node[operator] (M1) {M};
        &\node[phase] (Pc25) {}; 
        &&\node[operator] (M2) {M};
        &\coordinate (end5); &[-0.2cm]\node (d1) {$\cdots$};
        \\
        \node (q6) {$a_1$:}; % LINE a1
        & &\node[phase] (Pa11) {}; &\node[circlewc] (Ca11) {};  &\node[phase] (Pa12) {}; &
        &\node[phase] (Pa13) {}; &\node[circlewc] (Ca12) {}; &\node[phase] (Pa14) {}; &
        &\node[circlewc] (Ca13) {}; 
        &\node[operator] (M1) {M};
        &\node[phase] (Pa15) {};
        &\node[operator] (M2) {M};
        &\node[circlewc] (Ca14) {};
        &\node[operator] (M3) {M};
        &\node[phase] (Pa16) {};
        &&&&&\node[operator] (M4) {M};
        &\coordinate (end6); &[-0.2cm]\node (d1) {$\cdots$};
        \\ \node (q7) {$a_2$:}; % LINE a2
        &\node[phase] (Pa21) {}; &\node[circlewc] (Ca21) {}; &
        &\node[circlewc] (Ca22) {};
        &\node[phase] (Pa22) {}; 
        &\node[circlewc] (Ca23) {}; &
        &\node[circlewc] (Ca24) {}; &&&&&&&&&&& 
        &&
        &\coordinate (end7); &[0.2cm]\node (d1) {$\cdots$};
        \\
        &&&&& &&& &\node (down2) {}; &&&&&&& &\node (down3) {}; && &\node (down4) {}; % bottom line
        \\};
        \begin{pgfonlayer}{background} 
        \draw[thick] (q1) -- (end1) (q2) -- (end2) (q3) -- (end3) (q4) -- (end4) (q5) -- (end5) (q6) -- (end6) (q7) -- (end7)
        (Pc11) -- (Ca11)
        (P31) -- (Ca21) (Pa21) -- (C11)
        (P32) -- (Ca22) (Pa22) -- (C12)
        (Pc12) -- (Ca12)
        (P33) -- (Ca23)
        (P34) -- (Ca24)
        (Pc13) -- (Ca13)
        (Pa15) -- (C21)
        (Pc14) -- (Ca14)
        (Pa16) -- (C22)
        (C31) -- (Pc25);
        \draw[thick, dotted] (up2) -- (down2) (up3) -- (down3) (up4) -- (down4)
        ; \end{pgfonlayer}
    \end{tikzpicture}
    \caption{2q-coin QW on a 8-node ring with native gate set $\textbf{G}(\text{max}(r) = 3)$;  \textbf{increment only}.}
\end{figure}

% max3 Lazy 16-node
\begin{figure}[h!]
    \centering
    \begin{tikzpicture}[thick]
        \tikzset{
        operator/.style = {draw,fill=white,minimum size=0.1em}, 
        operator2/.style = {draw,fill=white,minimum height=2cm, minimum width=1cm}, 
        phase/.style = {draw,fill,shape=circle,minimum size=5pt,inner sep=0pt}, 
        surround/.style = {fill=blue!10,thick,draw=black,rounded corners=2mm}, 
        cross/.style={path picture={\draw[thick,black](path picture bounding box.north) -- (path picture bounding box.south) (path picture bounding box.west) -- (path picture bounding box.east); }}, crossx/.style={path picture={ \draw[thick,black,inner sep=0pt] (path picture bounding box.south east) -- (path picture bounding box.north west) (path picture bounding box.south west) -- (path picture bounding box.north east); }}, 
        circlewc/.style={draw,circle,cross,minimum width=0.3 cm}, 
        meter/.style= {draw, fill=white, inner sep=5, rectangle, font=\vphantom{A}, minimum width=20, line width=.8, path picture={\draw[black] ([shift={(.1,.2)}]path picture bounding box.south west) to[bend left=30] ([shift={(-.1,.2)}]path picture bounding box.south east);\draw[black,-latex] ([shift={(0,.1)}]path picture bounding box.south) -- ([shift={(.2,-.1)}]path picture bounding box.north);}}, }
        \matrix[row sep=0.2cm, column sep=0.05cm] (circuit) {
        &&&&&&&&&&&& &\node (up1) {}; &&&&&&&&&&&& &\node (up2) {}; &&&&&&& &\node (up3) {}; && &\node (up4) {};
        \\\node (q1) {$x_1$:}; % LINE x1
        &\node[circlewc] (C11) {}; &&&&& &\node[circlewc] (C12) {}; &&&&&&&&&&&&&&&&&&&&&&&&&&&&&&& &\coordinate (end1); &[0.05cm]\node (d1) {$\cdots$}; %&[0.05cm]\node[meter] (meter) {};
        \\ \node (q2) {$x_2$:}; % LINE x2
        &\node[phase] (P21) {}; &&&&& &\node[phase] (P22) {}; &&&&&&& &\node[circlewc] (C21) {}; &&
        &&&&\node[circlewc] (C22) {}; &&&&&&&&&&&&&&&&&
        &\coordinate (end2); &[-0.2cm]\node (d2) {$\cdots$}; %&[-0.3cm]\node[meter] (meter) {};
        \\ \node (q3) {$x_3$:}; % LINE x3
        & &\node[phase] (P31) {}; &&& &\node[phase] (P32) {}; & &\node[phase] (P33) {}; &&&&\node[phase] (P34) {}; &&&\node[phase] (P35) {}; &&&&&
        &\node[phase] (P36) {}; &&&&
        &&& &\node[circlewc] (C21n) {}; &&&
        &\node[circlewc] (C22n) {}; &&&&&
        &\coordinate (end3); &[-0.2cm]\node (d2) {$\cdots$}; %&[-0.3cm]\node[meter] (meter) {};
        \\ \node (q3n) {$x_4$:}; % LINE x4 (3n)
        & & &\node[phase] (Px41) {}; & &\node[phase] (Px42) {}; &&& &\node[phase] (Px43) {}; &&\node[phase] (Px44) {}; &&&&&
        &\node[phase] (Px45) {}; &
        &\node[phase] (Px46) {}; &&&&\node[phase] (Px47) {}; &&\node[phase] (Px48) {};
        &&& &\node[phase] (P35n) {}; &&&
        &\node[phase] (P36n) {}; && &\node[circlewc] (C31n) {}; &&
        &\coordinate (end3n); &[-0.2cm]\node (d2) {$\cdots$}; %&[-0.3cm]\node[meter] (meter) {};
        \\ \node (q4) {$c_1$:}; % LINE c1
        &&& &\node[phase] (Pc11) {}; &&&
        &&&\node[phase] (Pc12) {}; &&&&&&&
        &\node[phase] (Pc13) {}; &
        &&&&&\node[phase] (Pc14) {}; &
        &
        &\node[phase] (Pc13n) {}; &&&
        &\node[phase] (Pc14n) {}; &&&&
        &\node[phase] (Pc15n) {}; &&
        &\coordinate (end4); &[-0.2cm]\node (d1) {$\cdots$};
        \\ \node (q5) {$c_2$:}; % LINE c2
        &&&&\node[phase] (P0) {}; &&&&&
        &\node[phase] (P0) {}; &&&&&&&
        &\node[phase] (P0) {}; &&&&&
        &\node[phase] (Pc24) {}; &
        &
        &\node[phase] (Pc23) {}; &&&
        &\node[phase] (Pc24) {}; &&&
        &\node[operator] (M1) {M};
        &\node[phase] (Pc25) {}; 
        &&\node[operator] (M2) {M}; 
        &\coordinate (end5); &[-0.2cm]\node (d1) {$\cdots$};
        \\
        \node (q6) {$a_1$:}; % LINE a1
        & &&\node[phase] (P0) {}; &\node[circlewc] (Ca11) {};  &\node[phase] (P0) {}; &&&&\node[phase] (P0) {}; &\node[circlewc] (Ca12) {}; &\node[phase] (P0) {}; &&&&
        &&\node[phase] (P0) {}; &\node[circlewc] (Ca13) {}; &\node[phase] (P0) {};
        &&&&\node[phase] (P0) {};
        &\node[circlewc] (Ca14) {};
        &\node[phase] (P0) {}; &
        &\node[circlewc] (Ca13n) {}; 
        &\node[operator] (M1) {M};
        &\node[phase] (Pa15n) {};
        &\node[operator] (M2) {M};
        &\node[circlewc] (Ca14n) {};
        &\node[operator] (M3) {M};
        &\node[phase] (Pa16n) {};
        &&&&&\node[operator] (M4) {M}; 
        &\coordinate (end6); &[-0.2cm]\node (d1) {$\cdots$};
        \\ \node (q7) {$a_2$:}; % LINE a2
        &&\node[phase] (P0) {}; &\node[circlewc] (Ca21N) {}; &
        &\node[circlewc] (Ca22N) {};
        &\node[phase] (P0) {}; &&\node[phase] (P0) {};
        &\node[circlewc] (Ca23N) {}; &&\node[circlewc] (Ca24N) {}; &\node[phase] (P0) {}; &
        &\node[operator] (M) {M};
        &\node[phase] (Pa15) {}; &\node[operator] (M) {M};
        &\node[circlewc] (Ca211N) {}; &
        &\node[circlewc] (Ca212N) {};
        &\node[operator] (M) {M}; &\node[phase] (P22) {};
        &\node[operator] (M) {M};
        &\node[circlewc] (Ca213N) {}; &
        &\node[circlewc] (Ca214N) {}; &&&&&&&&&&&&&
        &\coordinate (end7); &[0.2cm]\node (d1) {$\cdots$};
        \\ \node (q8) {$a_3$:}; % LINE a3
        &\node[phase] (Pa21) {}; &\node[circlewc] (Ca21) {}; &
        &&&\node[circlewc] (Ca22) {};
        &\node[phase] (Pa22) {};
        &\node[circlewc] (Ca23) {}; &&&
        &\node[circlewc] (Ca24) {}; &&&&& 
        &&&&&&&&&&&&&&&&&&&&&
        &\coordinate (end8); &[0.2cm]\node (d1) {$\cdots$};
        \\
        &&&&&&&&&&&&& \node (down1) {}; &&&&&&&&&&&& &\node (down2) {}; &&&&&&& &\node (down3) {}; && &\node (down4) {};
        \\};
        \begin{pgfonlayer}{background} 
        \draw[thick] (q1) -- (end1) (q2) -- (end2) (q3) -- (end3) (q4) -- (end4) (q5) -- (end5) (q6) -- (end6) (q7) -- (end7) (q8) -- (end8) (q3n) -- (end3n)
        (C22) -- (P22)
        (Pc11) -- (Ca11)
        (P31) -- (Ca21) (Pa21) -- (C11)
        (P32) -- (Ca22) (Pa22) -- (C12)
        (Pc12) -- (Ca12)
        (P33) -- (Ca23)
        (P34) -- (Ca24)
        (Pc13) -- (Ca13)
        (Pa15) -- (C21)
        (Pc14) -- (Ca14)
        %(Pa16) -- (C22)
        %(C31) -- (Pc25)
        (C31n) -- (Pc25)
        (C21n) -- (Pa15n) (C22n) -- (Pa16n)
        (Ca13n) -- (Pc13n) (Ca14n) -- (Pc14n)
        (Ca21N) -- (Px41) (Ca22N) -- (Px42)
        (Ca23N) -- (Px43) (Ca24N) -- (Px44)
        (Ca211N) -- (Px45) (Ca212N) -- (Px46) (Ca213N) -- (Px47) (Ca214N) -- (Px48)
        ;
        \draw[thick,dotted] (up1) -- (down1) (up2) -- (down2) (up3) -- (down3) (up4) -- (down4)
        ; \end{pgfonlayer}
    \end{tikzpicture}
    \caption{2q-coin QW on a 16-node ring with native gate set $\textbf{G}(\text{max}(r) = 3)$;  \textbf{increment only}.}
\end{figure}

%%%%%%%%%%%%%%%%%%%%%%%%%%%%%%%%%%%%%%%%%%%%%%%%%%%%%%%%%%%%%%%%%%%%%%
\section{QW circuits with native gates of maximum rank 4}\label{app:circ4}
%%%%%%%%%%%%%%%%%%%%%%%%%%%%%%%%%%
\subsection{1q-coin QWs}
% max4 NL 8-node
\begin{figure}[h!]
    \centering
    \begin{tikzpicture}[thick]
        \tikzset{
        operator/.style = {draw,fill=white,minimum size=0.1em}, 
        operator2/.style = {draw,fill=white,minimum height=2cm, minimum width=1cm}, 
        phase/.style = {draw,fill,shape=circle,minimum size=5pt,inner sep=0pt}, 
        surround/.style = {fill=blue!10,thick,draw=black,rounded corners=2mm}, 
        cross/.style={path picture={\draw[thick,black](path picture bounding box.north) -- (path picture bounding box.south) (path picture bounding box.west) -- (path picture bounding box.east); }}, crossx/.style={path picture={ \draw[thick,black,inner sep=0pt] (path picture bounding box.south east) -- (path picture bounding box.north west) (path picture bounding box.south west) -- (path picture bounding box.north east); }}, 
        circlewc/.style={draw,circle,cross,minimum width=0.3 cm}, 
        meter/.style= {draw, fill=white, inner sep=5, rectangle, font=\vphantom{A}, minimum width=20, line width=.8, path picture={\draw[black] ([shift={(.1,.2)}]path picture bounding box.south west) to[bend left=30] ([shift={(-.1,.2)}]path picture bounding box.south east);\draw[black,-latex] ([shift={(0,.1)}]path picture bounding box.south) -- ([shift={(.2,-.1)}]path picture bounding box.north);}}, }
        \matrix[row sep=0.2cm, column sep=0.35cm] (circuit) {
        \node (q1) {$x_1$: $\ket{0}$}; % LINE p1
        & &\node[circlewc] (C11) {}; & & & &\node[circlewc] (C12) {}; &&&&&&\coordinate (end1); &[-0.2cm]\node (d1) {$\cdots$}; &[-0.3cm]\node[meter] (meter) {};
        \\ \node (q2) {$x_2$: $\ket{0}$}; % LINE p2
        & &\node[phase] (P0) {}; &\node[circlewc] (C21) {}; & &\node[operator] (X2) {X}; &\node[phase] (P0) {}; &\node[operator] (X21) {X}; &\node[circlewc] (C22) {}; &&&& \coordinate (end2); &[-0.2cm]\node (d1) {$\cdots$}; &[-0.3cm]\node[meter] (meter) {};
        \\ \node (q3) {$x_3$: $\ket{0}$}; % LINE p3
        & &\node[phase] (P0) {}; &\node[phase] (P0) {}; &\node[circlewc] (C31) {}; &\node[operator] (X3) {X}; &\node[phase] (P0) {}; &&\node[phase] (P0) {}; &\node[operator] (X31) {X}; &\node[circlewc] (C32) {}; && \coordinate (end3); &[-0.2cm]\node (d1) {$\cdots$}; &[-0.3cm]\node[meter] (meter) {};
        \\ \node (q4) {$c_1$: $\ket{0}$}; % LINE c1
        &\node[operator] (H11) {$\text{R}_y(\theta_t)$}; &\node[phase] (P1) {}; &\node[phase] (P2) {}; &\node[phase] (P3) {}; &\node[operator] (X1) {X}; &\node[phase] (P4) {}; &&\node[phase] (P5) {}; &&\node[phase] (P6) {}; &\node[operator] (X11) {X}; & \coordinate (end4); &[-0.2cm]\node (d1) {$\cdots$};
        \\ };
        \begin{pgfonlayer}{background} 
        \draw[thick] (q1) -- (end1) (q2) -- (end2) (q3) -- (end3) (q4) -- (end4) (P1) -- (C11) (P2) -- (C21) (P3) -- (C31) (P4) -- (C12) (P5) -- (C22) (P6) -- (C32); \end{pgfonlayer}
    \end{tikzpicture}
    \caption{1q-coin QW on a 8-node ring with native gate set $\textbf{G}(\text{max}(r) = 4)$.}
\end{figure}

% max4 NL 16-node
\begin{figure}[h!]
    \centering
        \begin{tikzpicture}[thick]
        \tikzset{
        operator/.style = {draw,fill=white,minimum size=0.1em}, 
        operator2/.style = {draw,fill=white,minimum height=2cm, minimum width=1cm}, 
        phase/.style = {draw,fill,shape=circle,minimum size=5pt,inner sep=0pt}, 
        surround/.style = {fill=blue!10,thick,draw=black,rounded corners=2mm}, 
        cross/.style={path picture={\draw[thick,black](path picture bounding box.north) -- (path picture bounding box.south) (path picture bounding box.west) -- (path picture bounding box.east); }}, crossx/.style={path picture={ \draw[thick,black,inner sep=0pt] (path picture bounding box.south east) -- (path picture bounding box.north west) (path picture bounding box.south west) -- (path picture bounding box.north east); }}, 
        circlewc/.style={draw,circle,cross,minimum width=0.3 cm}, 
        meter/.style= {draw, fill=white, inner sep=5, rectangle, font=\vphantom{A}, minimum width=20, line width=.8, path picture={\draw[black] ([shift={(.1,.2)}]path picture bounding box.south west) to[bend left=30] ([shift={(-.1,.2)}]path picture bounding box.south east);\draw[black,-latex] ([shift={(0,.1)}]path picture bounding box.south) -- ([shift={(.2,-.1)}]path picture bounding box.north);}}, }
        \matrix[row sep=0.2cm, column sep=0.35cm] (circuit) {
        &&&& &\node (up1) {}; && &\node (up2) {}; & &\node (up3) {}; & &\node (up4) {}; % top line
        \\ \node (q1) {$x_1$:}; % LINE x1
        & &\node[circlewc] (Ctop1) {}; &&\node[circlewc] (Ctop2) {}; &&&&&&&&& &\coordinate (end1); &[-0.2cm]\node (d1) {$\cdots$}; %&[-0.3cm]\node[meter] (meter) {};
        \\ \node (q2) {$x_2$:}; % LINE x2
        &&\node[phase] (P0) {}; &&\node[phase] (P0) {}; &&&\node[circlewc] (C21) {}; &&&&&& &\coordinate (end2); &[-0.2cm]\node (d1) {$\cdots$}; %&[-0.3cm]\node[meter] (meter) {};
        \\ \node (q3) {$x_3$:}; % LINE x3
         &&\node[phase] (P0) {}; &&\node[phase] (P0) {}; &&&\node[phase] (P0) {}; &&\node[circlewc] (C31) {}; &&&& &\coordinate (end3); &[-0.2cm]\node (d1) {$\cdots$}; %&[-0.3cm]\node[meter] (meter) {};
        \\ \node (q4) {$x_4$:}; % LINE x4
        &\node[phase] (Pc1) {}; &&\node[phase] (Pc2) {}; &&&\node[operator] (M1) {M}; &\node[phase] (P0) {}; &&\node[phase] (P0) {}; &&\node[circlewc] (Cend) {}; &&\node[operator] (M1) {M}; &\coordinate (end4); &[-0.2cm]\node (d1) {$\cdots$};
        \\ \node (q5) {$c_1$:}; % LINE c1
        &\node[phase] (P1) {}; &&\node[phase] (P1) {}; &&&\node[operator] (M1) {M}; &\node[phase] (P2) {}; &&\node[phase] (P3) {}; &&\node[phase] (Pend) {}; &&\node[operator] (M1) {M}; &\coordinate (end5); &[-0.2cm]\node (d1) {$\cdots$};
        \\ \node (q6) {$a_1$:}; % LINE a1
        &\node[circlewc] (Ca1) {}; &\node[phase] (Pa1) {};
        &\node[circlewc] (Ca2) {}; &\node[phase] (Pa2) {};
        &&&&&&&&& &\coordinate (end6); &[-0.2cm]\node (d1) {$\cdots$};
        \\ 
        \\ &&&& &\node (down1) {}; && &\node (down2) {}; & &\node (down3) {}; & &\node (down4) {};% bottom line
        \\};
        \begin{pgfonlayer}{background} 
        \draw[thick] (q1) -- (end1) (q2) -- (end2) (q3) -- (end3) (q4) -- (end4) (q5) -- (end5) (q6) -- (end6) (P2) -- (C21) (P3) -- (C31) (Ctop1) -- (Pa1) (Ctop2) -- (Pa2) (Ca1) -- (Pc1) (Ca2) -- (Pc2) (Cend) -- (Pend);
        \draw[thick, dotted] (up1) -- (down1) (up2) -- (down2) (up3) -- (down3) (up4) -- (down4)
        ; \end{pgfonlayer}
    \end{tikzpicture}
    \caption{1q-coin QW on a 16-node ring with native gate set $\textbf{G}(\text{max}(r) = 4)$;  \textbf{increment only}.}
\end{figure}

%%%%%%%%%%%%%%%%%%%%%%%%%%%%%%%%%%%%%%
\newpage
\subsection{2q-coin QWs}

% max4 Lazy 4-node
\begin{figure}[h!]
    \centering
    \begin{tikzpicture}[thick]
        \tikzset{
        operator/.style = {draw,fill=white,minimum size=0.1em}, 
        operator2/.style = {draw,fill=white,minimum height=2cm, minimum width=1cm}, 
        phase/.style = {draw,fill,shape=circle,minimum size=5pt,inner sep=0pt}, 
        surround/.style = {fill=blue!10,thick,draw=black,rounded corners=2mm}, 
        cross/.style={path picture={\draw[thick,black](path picture bounding box.north) -- (path picture bounding box.south) (path picture bounding box.west) -- (path picture bounding box.east); }}, crossx/.style={path picture={ \draw[thick,black,inner sep=0pt] (path picture bounding box.south east) -- (path picture bounding box.north west) (path picture bounding box.south west) -- (path picture bounding box.north east); }}, 
        circlewc/.style={draw,circle,cross,minimum width=0.3 cm}, 
        meter/.style= {draw, fill=white, inner sep=5, rectangle, font=\vphantom{A}, minimum width=20, line width=.8, path picture={\draw[black] ([shift={(.1,.2)}]path picture bounding box.south west) to[bend left=30] ([shift={(-.1,.2)}]path picture bounding box.south east);\draw[black,-latex] ([shift={(0,.1)}]path picture bounding box.south) -- ([shift={(.2,-.1)}]path picture bounding box.north);}}, }
        \matrix[row sep=0.2cm, column sep=0.35cm] (circuit) {
        \node (q1) {$x_1$: $\ket{0}$}; % LINE p1
        & &\node[circlewc] (C1) {}; & & &\node[circlewc] (C12) {}; &&&&\coordinate (end1); &[-0.2cm]\node (d1) {$\cdots$}; &[-0.3cm]\node[meter] (meter) {};
        \\ \node (q2) {$x_2$: $\ket{0}$}; % LINE p2
        &&\node[phase] (P0) {}; &\node[circlewc] (C21) {}; &\node[operator] (X0) {X}; &\node[phase] (P0) {}; &\node[operator] (X0) {X}; &\node[circlewc] (C22) {}; & &\coordinate (end2); &[-0.2cm]\node (d1) {$\cdots$}; &[-0.3cm]\node[meter] (meter) {};
        \\ \node (q3) {$c_1$: $\ket{0}$}; % LINE c1
        &\node[operator] (H11) {$\text{R}_y(\theta_t)$}; &\node[phase] (P11) {}; &\node[phase] (P12) {}; &\node[operator] (X1) {X}; &\node[phase] (P13) {}; &&\node[phase] (P13) {}; &\node[operator] (X2) {X}; & \coordinate (end3); &[-0.2cm]\node (d1) {$\cdots$};
        \\ \node (q4) {$c_2$: $\ket{0}$}; % LINE c2
        &\node[operator] (H11) {$\text{R}_y(\phi_t)$}; &\node[phase] (P22) {}; &\node[phase] (P21) {}; & &\node[phase] (P26) {}; &&\node[phase] (P25) {}; &&\coordinate (end4); &[-0.2cm]\node (d1) {$\cdots$};
        \\};
        \begin{pgfonlayer}{background} 
        \draw[thick] (q1) -- (end1) (q2) -- (end2) (q3) -- (end3) (q4) -- (end4) (P21) -- (C21) (P22) -- (C1) (P25) -- (C22) (P26) -- (C12); \end{pgfonlayer}
    \end{tikzpicture}
    \caption{2q-coin QW on a 4-node ring with native gate set $\textbf{G}(\text{max}(r) = 4)$.}
\end{figure}

% max4 Lazy 8-node
\begin{figure}[h!]
    \centering
    \begin{tikzpicture}[thick]
        \tikzset{
        operator/.style = {draw,fill=white,minimum size=0.1em}, 
        operator2/.style = {draw,fill=white,minimum height=2cm, minimum width=1cm}, 
        phase/.style = {draw,fill,shape=circle,minimum size=5pt,inner sep=0pt}, 
        surround/.style = {fill=blue!10,thick,draw=black,rounded corners=2mm}, 
        cross/.style={path picture={\draw[thick,black](path picture bounding box.north) -- (path picture bounding box.south) (path picture bounding box.west) -- (path picture bounding box.east); }}, crossx/.style={path picture={ \draw[thick,black,inner sep=0pt] (path picture bounding box.south east) -- (path picture bounding box.north west) (path picture bounding box.south west) -- (path picture bounding box.north east); }}, 
        circlewc/.style={draw,circle,cross,minimum width=0.3 cm}, 
        meter/.style= {draw, fill=white, inner sep=5, rectangle, font=\vphantom{A}, minimum width=20, line width=.8, path picture={\draw[black] ([shift={(.1,.2)}]path picture bounding box.south west) to[bend left=30] ([shift={(-.1,.2)}]path picture bounding box.south east);\draw[black,-latex] ([shift={(0,.1)}]path picture bounding box.south) -- ([shift={(.2,-.1)}]path picture bounding box.north);}}, }
        \matrix[row sep=0.2cm, column sep=0.35cm] (circuit) {
        &&&& &\node (up1) {}; && &\node (up2) {}; & &\node (up3) {}; % top line
        \\ \node (q1) {$x_1$:}; % LINE x1
        & &\node[circlewc] (Ctop1) {}; &&\node[circlewc] (Ctop2) {}; &&&&&&& &\coordinate (end1); &[-0.2cm]\node (d1) {$\cdots$}; %&[-0.3cm]\node[meter] (meter) {};
        \\ \node (q2) {$x_2$:}; % LINE x2
        &&\node[phase] (P0) {}; &&\node[phase] (P0) {}; &&&\node[circlewc] (C21) {}; & &&& &\coordinate (end2); &[-0.2cm]\node (d1) {$\cdots$}; %&[-0.3cm]\node[meter] (meter) {};
        \\ \node (q3) {$x_3$:}; % LINE x3
         &&\node[phase] (P0) {}; &&\node[phase] (P0) {}; &&&\node[phase] (P0) {}; &&\node[circlewc] (C31) {}; && &\coordinate (end3); &[-0.2cm]\node (d1) {$\cdots$}; %&[-0.3cm]\node[meter] (meter) {};
        \\ \node (q4) {$c_1$:}; % LINE c1
        &\node[phase] (Pc1) {}; &&\node[phase] (Pc2) {}; &&&\node[operator] (M1) {M}; &\node[phase] (P0) {}; &&\node[phase] (P0) {}; &&\node[operator] (M1) {M}; &\coordinate (end4); &[-0.2cm]\node (d1) {$\cdots$};
        \\ \node (q5) {$c_2$:}; % LINE c2
        &\node[phase] (P1) {}; &&\node[phase] (P1) {}; &&&\node[operator] (M1) {M}; &\node[phase] (P2) {}; &&\node[phase] (P3) {}; &&\node[operator] (M1) {M}; &\coordinate (end5); &[-0.2cm]\node (d1) {$\cdots$};
        \\ \node (q6) {$a_1$:}; % LINE a1
        &\node[circlewc] (Ca1) {}; &\node[phase] (Pa1) {};
        &\node[circlewc] (Ca2) {}; &\node[phase] (Pa2) {};
        &&&&&&& &\coordinate (end6); &[-0.2cm]\node (d1) {$\cdots$};
        \\ 
        \\ &&&& &\node (down1) {}; && &\node (down2) {}; & &\node (down3) {}; % bottom line
        \\};
        \begin{pgfonlayer}{background} 
        \draw[thick] (q1) -- (end1) (q2) -- (end2) (q3) -- (end3) (q4) -- (end4) (q5) -- (end5) (q6) -- (end6) (P2) -- (C21) (P3) -- (C31) (Ctop1) -- (Pa1) (Ctop2) -- (Pa2) (Ca1) -- (Pc1) (Ca2) -- (Pc2);
        \draw[thick, dotted] (up1) -- (down1) (up2) -- (down2) (up3) -- (down3)
        ; \end{pgfonlayer}
    \end{tikzpicture}
    \caption{2q-coin QW on a 8-node ring with native gate set $\textbf{G}(\text{max}(r) = 4)$;  \textbf{increment only}.}
\end{figure}

% max4 lazy 16-node [TO DO]
\begin{figure}[h!]
    \centering
    \begin{tikzpicture}[thick]
        \tikzset{
        operator/.style = {draw,fill=white,minimum size=0.1em}, 
        operator2/.style = {draw,fill=white,minimum height=2cm, minimum width=1cm}, 
        phase/.style = {draw,fill,shape=circle,minimum size=5pt,inner sep=0pt}, 
        surround/.style = {fill=blue!10,thick,draw=black,rounded corners=2mm}, 
        cross/.style={path picture={\draw[thick,black](path picture bounding box.north) -- (path picture bounding box.south) (path picture bounding box.west) -- (path picture bounding box.east); }}, crossx/.style={path picture={ \draw[thick,black,inner sep=0pt] (path picture bounding box.south east) -- (path picture bounding box.north west) (path picture bounding box.south west) -- (path picture bounding box.north east); }}, 
        circlewc/.style={draw,circle,cross,minimum width=0.3 cm}, 
        meter/.style= {draw, fill=white, inner sep=5, rectangle, font=\vphantom{A}, minimum width=20, line width=.8, path picture={\draw[black] ([shift={(.1,.2)}]path picture bounding box.south west) to[bend left=30] ([shift={(-.1,.2)}]path picture bounding box.south east);\draw[black,-latex] ([shift={(0,.1)}]path picture bounding box.south) -- ([shift={(.2,-.1)}]path picture bounding box.north);}}, }
        \matrix[row sep=0.2cm, column sep=0.35cm] (circuit) {
        &&&& &\node (up0) {}; &&&& &\node (up1) {}; && &\node (up2) {}; & &\node (up3) {}; % top line
        \\ \node (q0) {$x_1$:}; % LINE x1
        & &\node[circlewc] (Ctoptop1) {}; &&\node[circlewc] (Ctoptop2) {}; &&\node[operator] (M1) {M}; &&&&&&&&&&\node[operator] (M1) {M}; &\coordinate (end0); &[-0.2cm]\node (d1) {$\cdots$};
        \\ \node (q1) {$x_2$:}; % LINE x2
        & &\node[phase] (p0) {}; &&\node[phase] (p0) {}; &
        & &\node[circlewc] (Ctop1) {}; &&\node[circlewc] (Ctop2) {}; &&&&&&& &\coordinate (end1); &[-0.2cm]\node (d1) {$\cdots$}; %&[-0.3cm]\node[meter] (meter) {};
        \\ \node (q2) {$x_3$:}; % LINE x3
        &&\node[phase] (P0) {}; &&\node[phase] (P0) {}; &
        &&\node[phase] (P0) {}; &&\node[phase] (P0) {}; &&&\node[circlewc] (C21) {}; & &&& &\coordinate (end2); &[-0.2cm]\node (d1) {$\cdots$}; %&[-0.3cm]\node[meter] (meter) {};
        \\ \node (q3) {$x_4$:}; % LINE x4
        &\node[phase] (Px41) {}; &&\node[phase] (Px42) {}; &&
        &&\node[phase] (P0) {}; &&\node[phase] (P0) {}; &&&\node[phase] (P0) {}; &&\node[circlewc] (C31) {}; && &\coordinate (end3); &[-0.2cm]\node (d1) {$\cdots$}; %&[-0.3cm]\node[meter] (meter) {};
        \\ \node (q4) {$c_1$:}; % LINE c1
        &\node[phase] (Pc1) {}; &&\node[phase] (Pc2) {}; &&
        &\node[phase] (Pc1) {}; &&\node[phase] (Pc2) {}; &&&\node[operator] (M1) {M}; &\node[phase] (P0) {}; &&\node[phase] (P0) {}; &&\node[operator] (M1) {M}; &\coordinate (end4); &[-0.2cm]\node (d1) {$\cdots$};
        \\ \node (q5) {$c_2$:}; % LINE c2
        &\node[phase] (P1) {}; &&\node[phase] (P1) {}; &&
        &\node[phase] (P1) {}; &&\node[phase] (P1) {}; &&&\node[operator] (M1) {M}; &\node[phase] (P2) {}; &&\node[phase] (P3) {}; &&\node[operator] (M1) {M}; &\coordinate (end5); &[-0.2cm]\node (d1) {$\cdots$};
        \\ \node (q6) {$a_1$:}; % LINE a1
        &\node[circlewc] (Ca1a) {}; &\node[phase] (Pa1a) {};
        &\node[circlewc] (Ca2a) {}; &\node[phase] (Pa2a) {}; &
        &\node[circlewc] (Ca1) {}; &\node[phase] (Pa1) {};
        &\node[circlewc] (Ca2) {}; &\node[phase] (Pa2) {};
        &&&&&&& &\coordinate (end6); &[-0.2cm]\node (d1) {$\cdots$};
        \\ 
        \\ &&&& &\node (down0) {}; &&&& &\node (down1) {}; && &\node (down2) {}; & &\node (down3) {}; % bottom line
        \\};
        \begin{pgfonlayer}{background} 
        \draw[thick] (q0) -- (end0) (q1) -- (end1) (q2) -- (end2) (q3) -- (end3) (q4) -- (end4) (q5) -- (end5) (q6) -- (end6) (P2) -- (C21) (P3) -- (C31) (Ctop1) -- (Pa1) (Ctop2) -- (Pa2) (Ca1) -- (Pc1) (Ca2) -- (Pc2)
        (Ctoptop1) -- (Pa1a) (Ctoptop2) -- (Pa2a)
        (Ca1a) -- (Px41) (Ca2a) -- (Px42)
        ;
        \draw[thick, dotted] (up1) -- (down1) (up2) -- (down2) (up3) -- (down3) (up0) -- (down0)
        ; \end{pgfonlayer}
    \end{tikzpicture}
    \caption{2q-coin QW on a 16-node ring with native gate set $\textbf{G}(\text{max}(r) = 4)$;  \textbf{increment only}.}
\end{figure}